\documentclass[runningheads]{llncs}
\usepackage{graphicx}
\usepackage{comment}
\usepackage{amsmath,amssymb} 
\usepackage{color,url}

\usepackage{ulem}
\usepackage{xcolor}
\usepackage{subfigure}
\usepackage{booktabs, array, multirow, threeparttable}

\usepackage[width=122mm,left=12mm,paperwidth=146mm,height=193mm,top=12mm,paperheight=217mm]{geometry}

\begin{document}
	\pagestyle{headings}
	\mainmatter
	\def\ECCVSubNumber{2544}  
	
	\title{Early Exit or Not: Resource-Efficient Blind Quality Enhancement for Compressed Images}

	\titlerunning{RBQE}
	%
	\author{
		Qunliang Xing \and
		Mai Xu \and
		Tianyi Li \and
		Zhenyu Guan}%
	\authorrunning{Q. Xing et al.}
	%
	\institute{Beihang University, Beijing, China\\
	\email{\{xingql,maixu,tianyili,guanzhenyu\}@buaa.edu.cn}}
	
	\maketitle
	
	\begin{abstract}
		Lossy image compression is pervasively conducted to save communication bandwidth, resulting in undesirable compression artifacts. Recently, extensive approaches have been proposed to reduce image compression artifacts at the decoder side; however, they require a series of architecture-identical models to process images with different quality, which are inefficient and resource-consuming. Besides, it is common in practice that compressed images are with unknown quality and it is intractable for existing approaches to select a suitable model for blind quality enhancement. In this paper, we propose a resource-efficient blind quality enhancement (RBQE) approach for compressed images. Specifically, our approach blindly and progressively enhances the quality of compressed images through a dynamic deep neural network (DNN), in which an early-exit strategy is embedded. Then, our approach can automatically decide to terminate or continue enhancement according to the assessed quality of enhanced images. Consequently, slight artifacts can be removed in a simpler and faster process, while the severe artifacts can be further removed in a more elaborate process. Extensive experiments demonstrate that our RBQE approach achieves state-of-the-art performance in terms of both blind quality enhancement and resource efficiency.
		The code is available at \textit{https://github.com/RyanXingQL/RBQE}.
		\keywords{Blind quality enhancement $\cdot$ Compressed images $\cdot$ Resource-efficient $\cdot$ Early-exit} 
	\end{abstract}

	\section{Introduction}
	
	We are embracing an era of visual data explosion.\footnotetext[1]{\textit{Accepted by ECCV 2020. DOI: 10.1007/978-3-030-58517-4\_17.}}\footnotetext[2]{\textit{Corresponding author: Mai Xu.}}
	According to Cisco mobile traffic forecast~\cite{Cisco}, the amount of mobile visual data is predicted to grow nearly 10-fold from 2017 to 2022.
	To overcome the bandwidth-hungry bottleneck caused by a deluge of visual data, lossy image compression, such as JPEG~\cite{125072}, JPEG 2000~\cite{marcellin2000overview} and HEVC-MSP~\cite{sullivan2012overview}, has been pervasively used.
	However, compressed images inevitably suffer from compression artifacts, such as blocky effects, ringing effects and blurring, which severely degrade the Quality of Experience (QoE)~\cite{seshadrinathan2010study,tan2015video} and the performance of high-level vision tasks~\cite{hennings2008simultaneous,zhang2011close}.
	
	For enhancing the quality of compressed images, many approaches~\cite{dong2015compression,guo2016building,wang2017novel,li2017efficient,yang2018enhancing,he2018enhancing,8855019} have been proposed.
	Their basic idea is that one model needs to be trained for enhancing compressed images with similar quality reflected by a particular value of Quantization Parameter (QP)~\cite{sullivan2012overview}, and then a series of architecture-identical models need to be trained for enhancing compressed images with different quality.
	For example, \cite{wang2017novel,yang2018enhancing,8855019} train 5 deep models to handle compressed images with QP $= 22$, $27$, $32$, $37$ and $42$.
	There are three main drawbacks to these approaches.
	(1) QP cannot faithfully reflect image quality, and thus it is intractable to manually select a suitable model based on QP value.
	(2) These approaches consume large computational resources during the training stage since many architecture-identical models need to be trained.
	(3) Compressed images with different quality are enhanced with the same computational complexity, such that these approaches impose excessive computational costs on ``easy'' samples (high-quality compressed images) but lack sufficient computation on ``hard'' samples (low-quality compressed images).
	Intuitively, the quality enhancement of images with different quality can be partly shared in a single framework, such that the joint computational costs can be reduced.
	More importantly, slight artifacts should be removed in a simpler and faster process, while the severe artifacts need to be further removed through a more elaborate process.
	Therefore, an ideal framework should automatically conduct a simple or elaborate enhancement process by distinguishing ``easy'' and ``hard'' samples, as a blind quality enhancement task.

	\begin{figure}[t]
		\centering
		\includegraphics[width=1\linewidth]{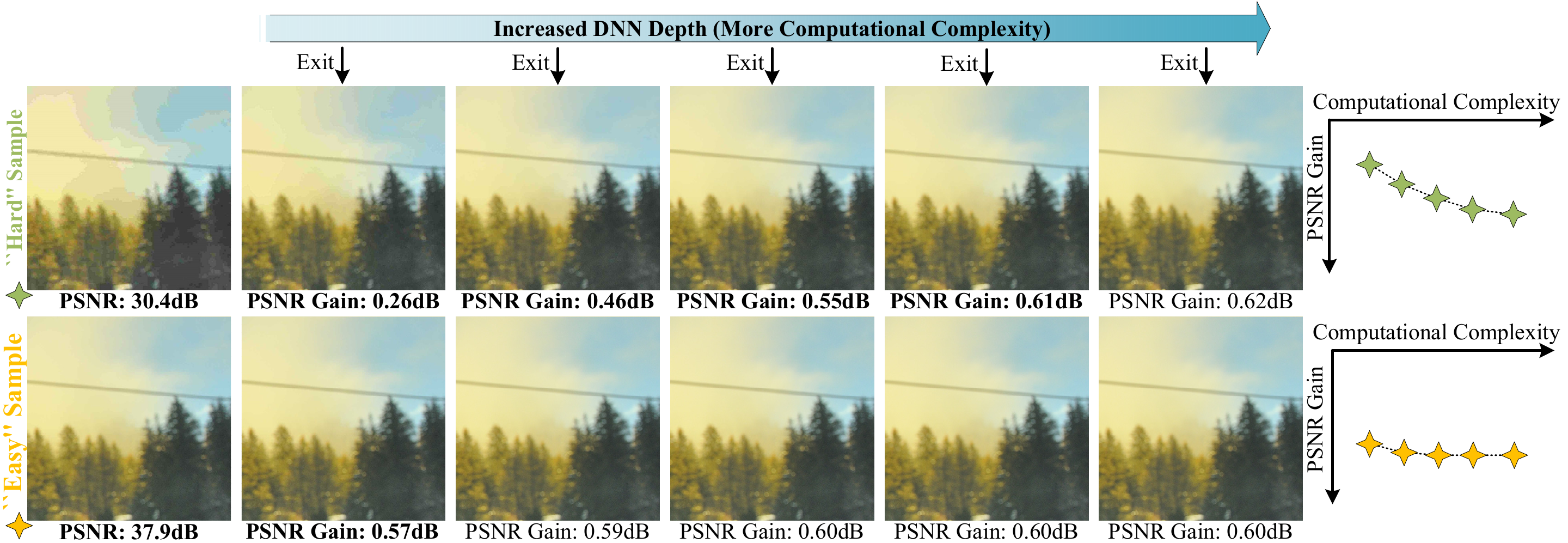}
		\caption{
			Examples of quality enhancement on ``easy'' and ``hard'' samples, along with increased computational complexity.
		}
		\label{fig:fig1}
	\end{figure}
	
	In this paper, we propose a resource-efficient blind quality enhancement (RBQE) approach for compressed images.
	Specifically, we first prove that there exist ``easy''/``hard'' samples for quality enhancement on compressed images.
	We demonstrate that ``easy'' samples are those with slight compression artifacts, while ``hard'' samples are those with severe artifacts.
	Then, a novel dynamic deep neural network (DNN) is designed, which progressively enhances the quality of compressed image, assesses the enhanced image quality, and automatically decides whether to terminate (early exit) or continue the enhancement.
	The quality assessment and early-exit decision are managed by a Tchebichef moments-based Image Quality Assessment Module (IQAM), which is strongly sensitive to compression artifacts.
	Finally, our RBQE approach can perform ``easy to hard'' quality enhancement in an end-to-end manner.
	This way, images with slight compression artifacts can be simply and rapidly enhanced, while those with severe artifacts need to be further enhanced.
	Some examples are shown in Fig.~\ref{fig:fig1}.
	Also, experimental results verify that our RBQE approach achieves state-of-the-art performance for blind quality enhancement in both efficiency and efficacy.
	
	To the best of knowledge, our approach is a first attempt to manage quality enhancement of compressed images in a resource-efficient manner. To sum up, the contributions are as follows:
	
	\begin{itemize}
		\item[(1)]{
			We prove that ``easy''/``hard'' samples exist in quality enhancement, as the theoretical foundation of our approach.
		}
		\item[(2)]{
			We propose the RBQE approach with a simple yet effective dynamic DNN architecture, which processes ``easy to hard'' paradigm for blind quality enhancement.
			
		}
		\item[(3)]{
			We develop a Tchebichef moments-based IQAM, workable for early-exit determination in our dynamic DNN structure.
		}
		
	\end{itemize}

	\section{Related Work}

	\begin{figure}[t]
		\centering
		\subfigure[]{
			\begin{minipage}[h]{1.0\linewidth}
				\centering
				\includegraphics[width=0.95\linewidth]{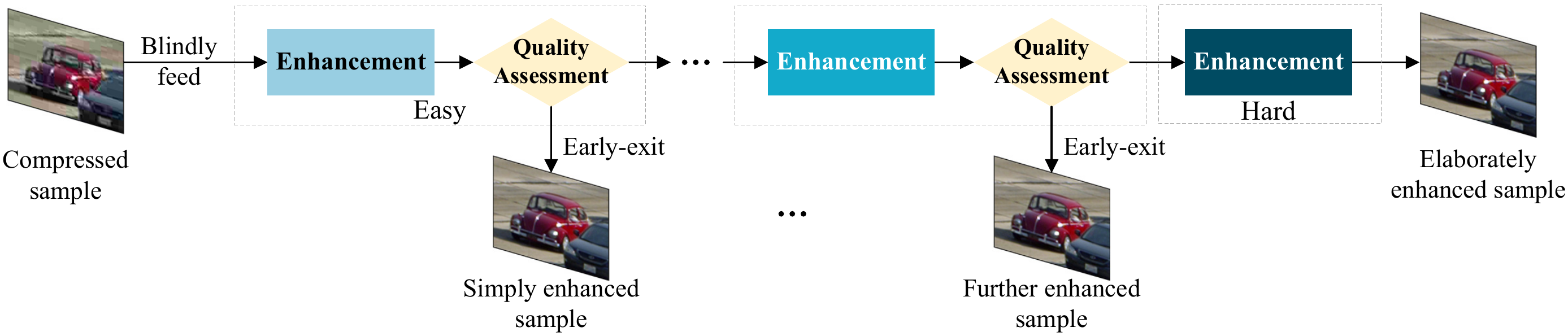}
		\end{minipage}}
		\subfigure[]{
			\begin{minipage}[h]{0.42\linewidth}
				\centering
				\includegraphics[width=1.0\linewidth]{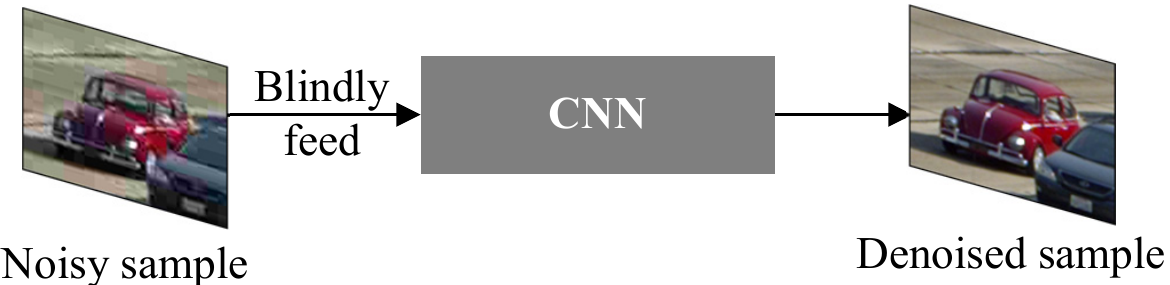}
		\end{minipage}}
		\subfigure[]{
			\begin{minipage}[h]{0.45\linewidth}
				\centering
				\includegraphics[width=1.0\linewidth]{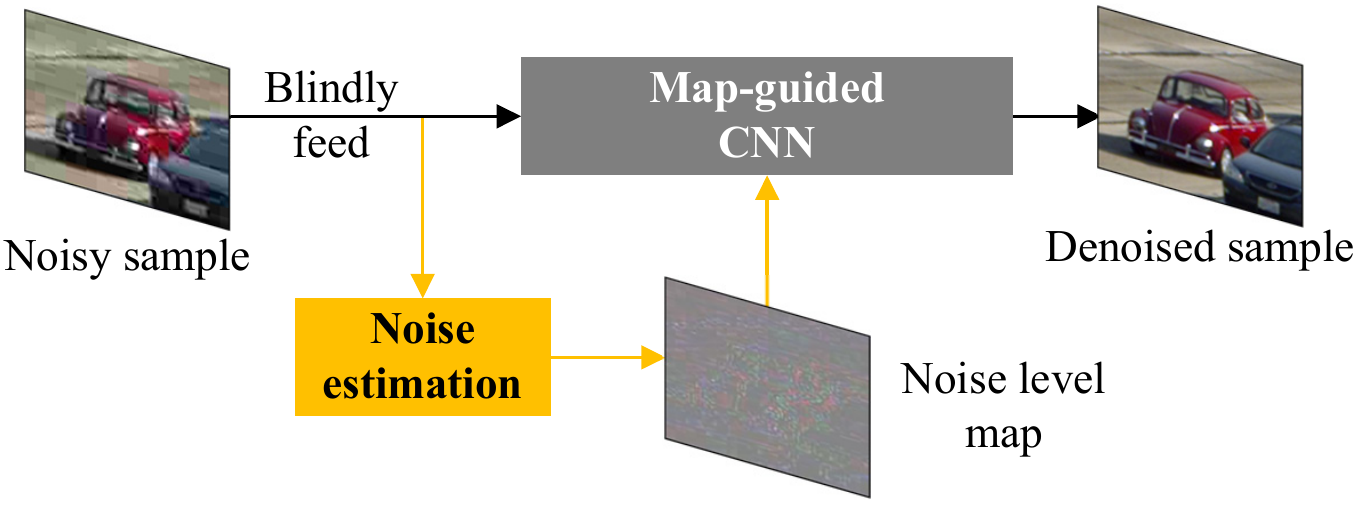}
		\end{minipage}}
		\caption{
			Proposed resource-efficient blind quality enhancement paradigm (a) vs. two traditional blind denoising paradigms (b) and (c).
			Our paradigm dynamically processes samples with early exits for ``easy'' samples, while traditional paradigms (b) and (c) statically process images with equal computational costs on both ``easy'' and ``hard'' samples.
		}
		\label{fig:related}
	\end{figure}
	
	\subsection{Quality Enhancement for Compressed Images}
	
	Due to the astonishing development of Convolutional Neural Networks (CNNs)~\cite{simonyan2014very} and large-scale image datasets~\cite{deng2009imagenet}, several CNN-based quality enhancement approaches have been successfully applied to JPEG-compressed images.
	Dong et al.~\cite{dong2015compression} proposed a shallow four-layer Artifacts Reduction Convolutional Neural Network (AR-CNN), which is the pioneer of CNN-based quality enhancement of JPEG-compressed images.
	Later, Deep Dual-Domain (D3) approach~\cite{wang2016d3} and Deep Dual-domain Convolutional neural Network (DDCN)~\cite{guo2016building} were proposed for JPEG artifacts removal, which are motivated by dual-domain sparse coding and utilize the quantization prior of JPEG compression.
	DnCNN~\cite{zhang2017beyond} is a milestone for reducing both Additive White Gaussian Noise (AWGN) and JPEG artifacts.
	It is a 20-layer deep network employing residual learning~\cite{he2016deep} and batch normalization~\cite{ioffe2015batch}, which can yield better results than Block-Matching and 3-D filtering (BM3D) approach~\cite{dabov2007image}.
	It also achieves blind denoising by mixing and sampling training data randomly with different levels of noise.
	
	Most recently, extensive works have been devoted to the latest video/image coding standard, HEVC/HEVC-MSP~\cite{sullivan2012overview,nguyen2012performance,cai2012lossy,nguyen2014objective,liu:hal-01876856,li2017closed}.
	Due to the elaborate coding strategies of HEVC, the approaches for JPEG-compressed images~\cite{dong2015compression,wang2016d3,guo2016building,zhang2017beyond}, especially those utilizing the prior of JPEG compression~\cite{wang2016d3,guo2016building}, cannot be directly used for quality enhancement of HEVC-compressed images.
	In fact, HEVC~\cite{sullivan2012overview} codec already incorporates the in-loop filters, which consist of Deblocking Filter (DF)~\cite{norkin2012hevc} and Sample Adaptive Offset (SAO) filter~\cite{fu2012sample}, to suppress blocky effects and ringing effects.
	However, these handcrafted filters are far from optimum, resulting in still visible artifacts in compressed images.
	To alleviate this issue, Wang et al.~\cite{wang2017novel} proposed the DCAD approach, which is the first attempt for CNN-based non-blind quality enhancement of HEVC-compressed images.
	Later, Yang et al.~\cite{yang2018enhancing} proposed a novel QE-CNN for quality enhancement of images compressed by HEVC-MSP.
	Unfortunately, they are all non-blind approaches, typically requiring QP information before quality enhancement.
	
	\subsection{Blind Denoising for Images}	
	
	In this section, we briefly review the CNN-based blind denoiser, as the closest field of blind quality enhancement of compressed images.
	The existing approaches for CNN-based blind denoising can be roughly summarized into two paradigms based on the mechanism of noise level estimation, as shown in Fig.~\ref{fig:related} (b) and (c).
	The first paradigm implicitly estimates the noise level.
	To achieve blind denoising, images with various levels of noise are mixed and randomly sampled during training~\cite{zhang2017beyond,tai2017memnet}.
	Unfortunately, the performance is always far from optimum, as stated in \cite{zhang2018ffdnet,guo2019toward}.
	It degrades severely when there is a mismatch of noise levels between training and test data.
	The second paradigm explicitly estimates the noise level.
	It sets a noise level estimation sub-net before a non-blind denoising sub-net.
	For example, \cite{guo2019toward} generates a noise level map to guide the subsequent non-blind denoising.
	This paradigm can always yield better results than the first paradigm, yet it is not suitable for quality enhancement of compressed artifacts, mainly due to two reasons.
	(1) The generated noise level map cannot well represent the level of compression artifacts.
	The compression artifacts are much more complex than generic noise since it is always assumed to be signal-independent and white~\cite{wang2016d3}.
	(2) Both ``easy'' and ``hard'' samples are processed in the same deep architecture consuming equal computational resources, resulting in low efficiency.
	In this paper, we provide a brand-new paradigm for image reconstruction (as shown in Fig.~\ref{fig:related} (a)) and exemplify it by our proposed RBQE on quality enhancement of compressed images.
	It is worth mentioning that our brand-new paradigm also has the potential for the blind denoising task.
	
	%
	
	\section{Proposed Approach}
	
	In this section, we propose our RBQE approach for blind quality enhancement.
	Specifically, we solve three challenging problems that are crucial to the resource-efficient paradigm of our approach.
	(1) Which samples are ``simple''/``hard'' in quality enhancement? (to be discussed in Sec.~\ref{sec:Motivation})
	(2) How to design a dynamic network for progressive enhancement? (to be discussed in Sec.~\ref{sec:network})
	(3) How to measure compression artifacts of enhanced compressed images for early exits? (to be discussed in Sec.~\ref{sec:IQAM})
	
	\subsection{Motivation}
	\label{sec:Motivation}
	
	Our RBQE approach is motivated by the following two propositions.
	\textbf{Proposition 1}: ``Easy'' samples (i.e., high-quality compressed images) can be simply enhanced, while ``hard'' samples (i.e., low-quality compressed images) should be further enhanced.
	\textbf{Proposition 2}: The quality enhancement process with different computational complexity can be jointly optimized in a single network through an ``easy to hard'' manner, rather than a ``hard to easy'' manner.
	
	\begin{figure}[t]
		\centering
		\subfigure[]{
			\begin{minipage}[h]{0.46\linewidth}
				\centering
				\includegraphics[width=1.0\linewidth]{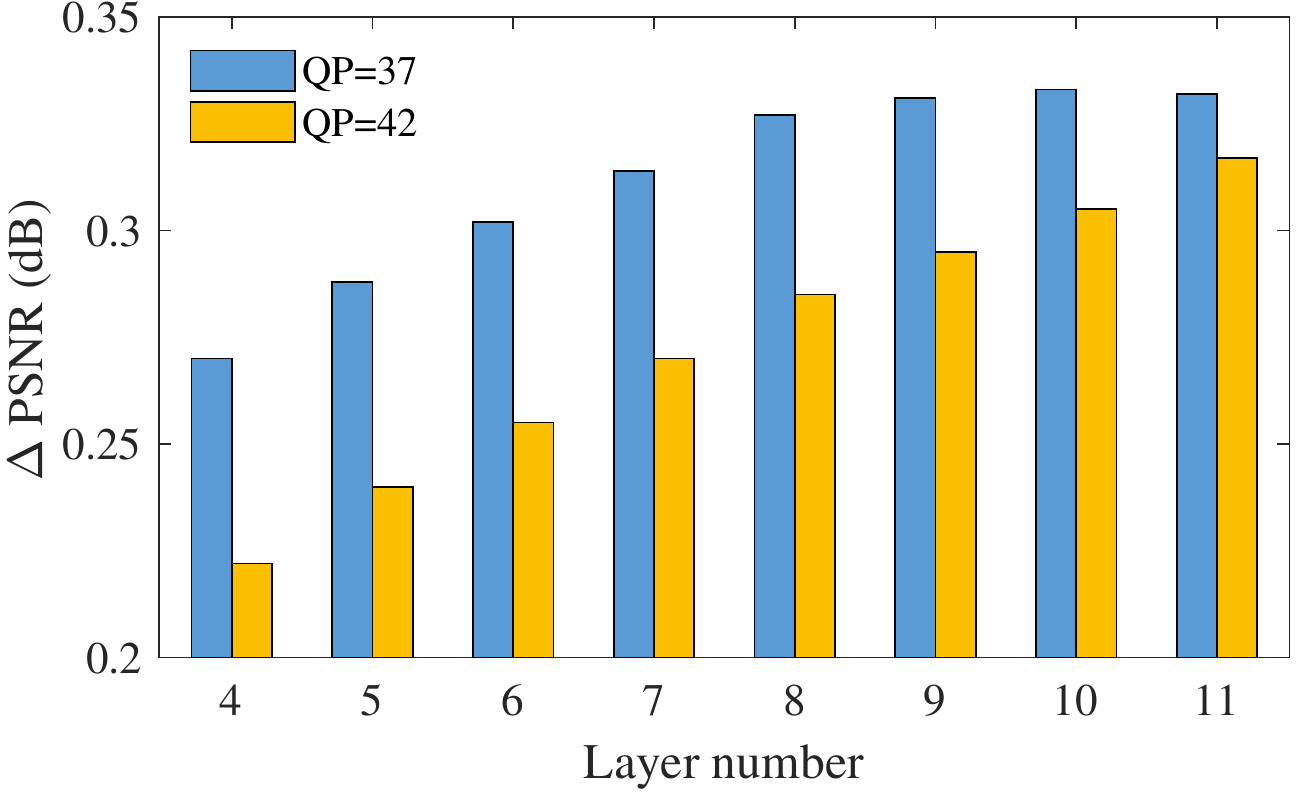}
			\end{minipage}
		}
		\subfigure[]{
			\begin{minipage}[h]{0.46\linewidth}
				\centering
				\includegraphics[width=1.0\linewidth]{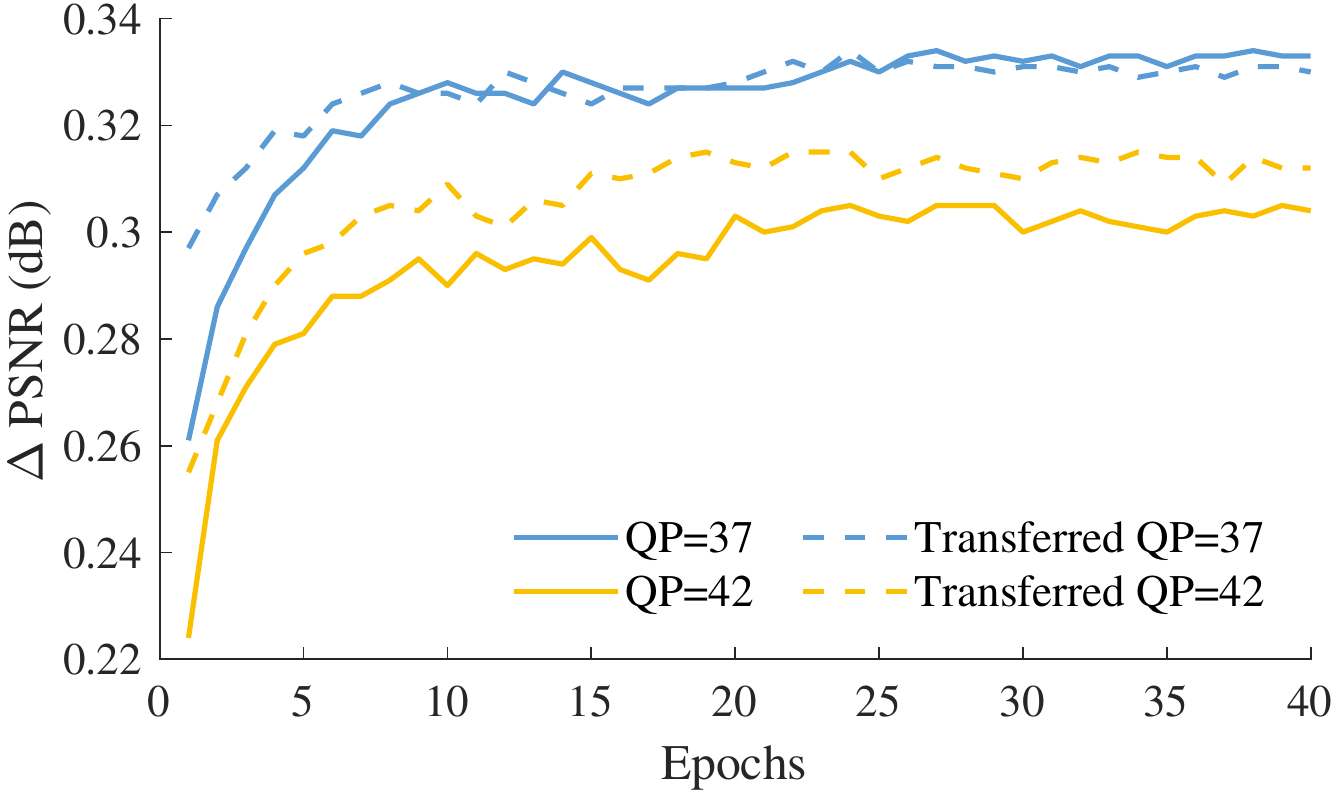}
			\end{minipage}
		}
		\caption{
			(a) Average improved peak signal-to-noise ratio ($\Delta$PSNR) of vanilla CNNs over the test set.
			(b) Average $\Delta$PSNR curves alongside increased epochs, for vanilla CNN models and their transferred models over the validation set during the training stage.
		}
		\label{fig:mot}
	\end{figure}
	
	\subsubsection{Proof of Proposition 1.}
	
	We construct a series of vanilla CNNs with different depths and feed them with ``easy'' and ``hard'' samples, respectively.
	Specifically, a series of vanilla CNNs with the layer number from 4 to 11 are constructed.
	Each layer includes $64 \times 3 \times 3$ filters, except for the last layer with $1 \times 3 \times 3$ filter.
	Beside, ReLU~\cite{nair2010rectified} activation and global residual learning~\cite{he2016deep} are adopted.
	The training, validation and test sets (including 400, 100 and 100 raw images, respectively) are randomly selected from Raw Image Database (RAISE)~\cite{dang2015raise} without overlapping.
	They are all compressed by HM16.5\footnote{HM16.5 is the latest HEVC reference software.} under intra-coding configuration~\cite{sullivan2012overview} with QP $=37$ and $42$ for obtaining ``easy'' and ``hard'' samples, respectively.
	Then, we train the vanilla CNNs with the ``easy'' samples, and then obtain converged models ``QP $=37$''.
	Similarly, we train the CNNs with the ``hard'' samples and then obtain converged models ``QP $=42$''.
	As shown in Fig.~\ref{fig:mot} (a), the performance of QP $=42$ models improves significantly with the increase of layer numbers, while the performance of QP $=37$ models gradually becomes saturated once the layer number excesses 9.
	Therefore, it is possible to enhance the ``easy'' samples with a simpler architecture and fewer computational resources, while further enhancing the ``hard'' samples in a more elaborate process.
	
	\subsubsection{Proof of Proposition 2.}
	
	The advantage of the ``easy to hard'' strategy has been pointed out in neuro-computation~\cite{gluck1993hippocampal}.
	Here, we investigate its efficacy on image quality enhancement through the experiments of transfer learning.
	If the filters learned from ``easy'' samples can be transferred to enhance ``hard'' samples more successfully than the opposite manner, then our proposition can be proved.
	Here, we construct 2 identical vanilla CNNs with 10 convolutional layers.
	The other settings conform to the above.
	We train these 2 models with the training sets of images compressed at QP $= 37$ and $42$, respectively, and accordingly these 2 models are called ``QP $=37$'' and ``QP $=42$''.
	After convergence, they exchange their parameters for the first 4 layers and restart training with their own training sets.
	Note that the exchanged parameters are frozen during the training stage.
	We name the model transferred from QP $=42$ to QP $=37$ as ``transferred QP $=37$'' and the model transferred from QP $=37$ to QP $=42$ as ``transferred QP $=42$''.
	Fig.~\ref{fig:mot} (b) shows the validation-epoch curves of the original 2 models and their transferred models.
	As shown in this figure, the transferred QP $=42$ model improves the performance of the QP $=42$ model, while the transferred QP $=37$ model slightly degrades the performance of the QP $=37$ model.
	Consequently, the joint simple and elaborate enhancement process should be conducted in an ``easy to hard'' manner rather than a ``hard to easy'' manner.
	Besides, the experimental results of Section~\ref{sec:exp} show that the simple and elaborate enhancement process can be jointly optimized in a single network.
	In summary, proposition 2 can be proved.
	The above propositions can be also validated by JPEG-compressed images, as detailed in the supplementary material.
	
	Given the above two propositions, we propose our RBQE approach for resource-efficient quality enhancement of compressed images in an ``easy to hard'' manner.
	
	\subsection{Dynamic DNN Architecture with Early-exit Strategy}
	\label{sec:network}
	
	\subsubsection{Notations.}
	
	In this section, we present the DNN architecture of the proposed RBQE approach for resource-efficient quality enhancement.
	We first introduce the notations for our RBQE approach.
	The input sample is denoted by $\mathbf{S}_{\text{in}}$.
	The convolutional layer is denoted by $C_{i,j}$, where $i$ denotes the level and $j$ denotes the index of the convolutional layer on the same level.
	In addition, $I$ is the total number of levels.
	Accordingly, the feature maps generated from $C_{i,j}$ are denoted by $\mathbf{F}_{i,j}$.
	The enhancement residuals are denoted by $\{\mathbf{R}_{j}\}_{j=2}^{I}$.
	Accordingly, the output enhanced samples are denoted by $\{\mathbf{S}_{\text{out},j}\}_{j=2}^{I}$.
	
	\subsubsection{Architecture.}
	
	\begin{figure}[t]
		\centering
		\includegraphics[width=1.0\linewidth]{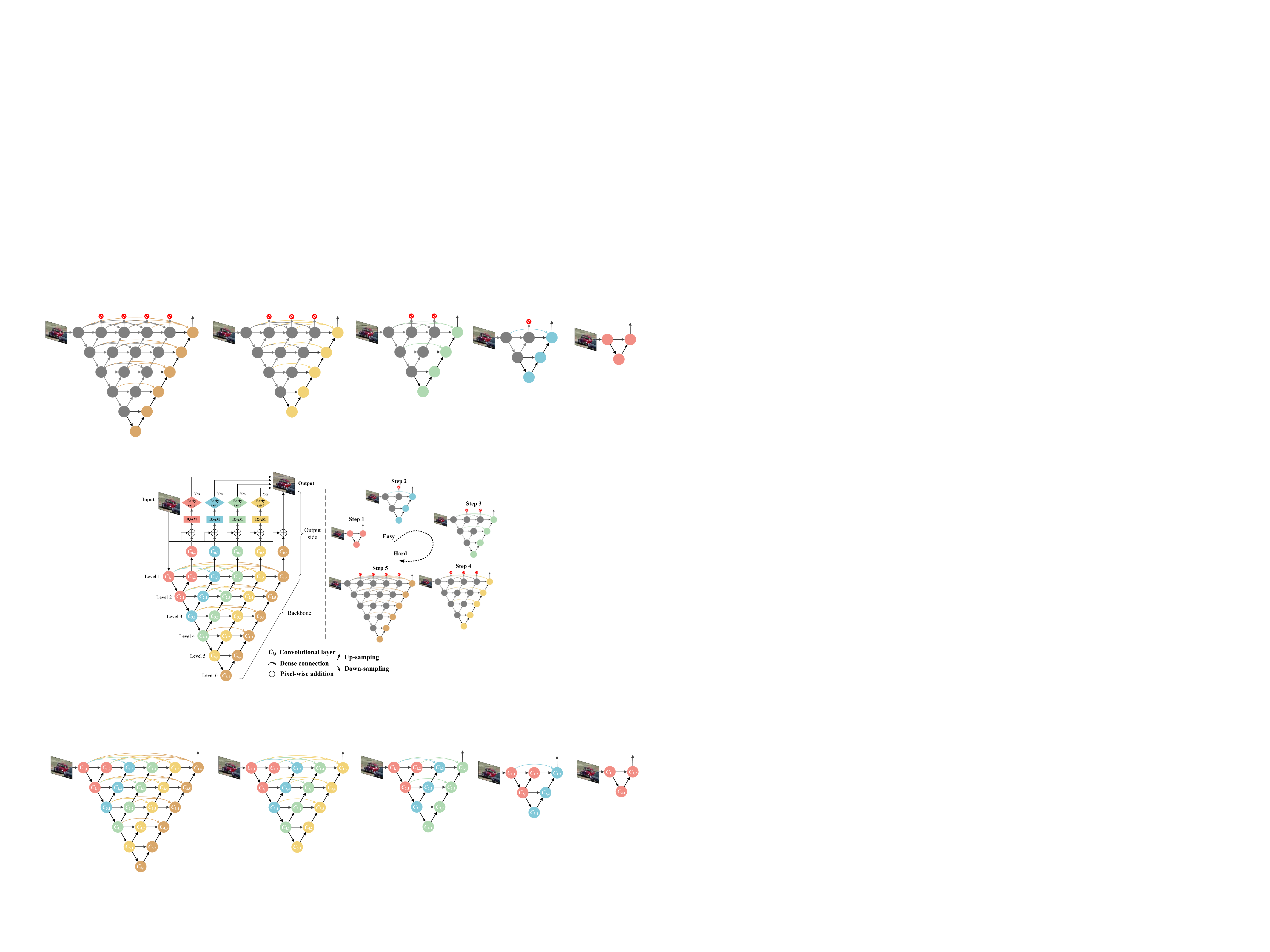}
		\caption{
			Dynamic DNN architecture and early-exit strategy of our RBQE approach.
			The computations of gray objects (arrays and circles) are accomplished in the previous step and inherited in the current step.
		}
		\label{fig:RBQE}
	\end{figure}
	
	To better illustrate the architecture of RBQE, we separate the backbone and the output side of RBQE, as shown in the left half of Fig.~\ref{fig:RBQE}.
	In this figure, we take RBQE with 6 levels as an example.
	The backbone of RBQE is a progressive UNet-based structure.
	Convolutional layers $C_{1,1}$ and $C_{2,1}$ can be seen as the encoding path of the smallest 2-level UNet, while $\{C_{i,1}\}_{i=1}^{6}$ are the encoding path of the largest 6-level UNet.
	Therefore, the backbone of RBQE can be considered as a compact combination of 5 different-level UNets.
	In the backbone of RBQE, the input sample is first fed into the convolutional layer $C_{1,1}$.
	After that, the feature maps generated by $C_{1,1}$ (i.e., $\mathbf{F}_{1,1}$) are progressively down-sampled and convoluted by $\{C_{i,1}, i = 2,3,...,6\}$.
	This way, we obtain feature maps $\{\mathbf{F}_{i,1}\}_{i=1}^{6}$ at 6 different levels, the size of which progressively becomes smaller from level 1 to 6.
	In accordance with the encoder-decoder architecture of the UNet approach, $\{\mathbf{F}_{i,1}\}_{i=1}^{6}$ are then progressively up-sampled and convoluted until level 1.
	Moreover, based on the progressive UNet structure, we adopt dense connections~\cite{huang2017densely} at each level.
	For example, at level 1, $\mathbf{F}_{1,1}$ are directly fed into the subsequent convolutional layers at the same level: $\{C_{1,j}\}_{j=2}^{6}$.
	The adoption of dense connection does not only encourage the reuse of encoded low-level fine-grained features by decoders, but also largely decreases the number of parameters, leading to a lightweight structure for RBQE.
	
	At the output side, the obtained feature maps $\{\mathbf{F}_{1,j}\}_{j=2}^{6}$ are further convoluted by independent convolutional layers $\{C_{0,j}\}_{j=2}^{6}$, respectively.
	In this step, we obtain the enhancement residuals: $\{\mathbf{R}_{j}\}_{j=2}^{6}$.
	For each residual $\mathbf{R}_{j}$, it is then added into the input sample $\mathbf{S}_{\text{in}}$ for calculating the enhanced image $\mathbf{S}_{\text{out},j}$:
	\begin{eqnarray}
	\mathbf{S}_{\text{out},j} = \mathbf{S}_{\text{in}} + \mathbf{R}_{j}.
	\end{eqnarray}
	To assess the quality of enhanced image $\mathbf{S}_{\text{out},j}$, we feed it into IQAM, which is to be presented in Section~\ref{sec:IQAM}.
	
	The backbone of RBQE is motivated by \cite{zhou2018unet++}, which extends the UNet architecture to a wide UNet for medical image segmentation.
	Here, we advance the wide UNet in the following aspects:
	(1) The wide UNet adopts deep supervision~\cite{lee2015deeply} directly for the feature maps $\{\mathbf{F}_{1,j}\}_{j=2}^{5}$.Here, we further process the output feature maps $\{\mathbf{F}_{1,j}\}_{j=2}^{I}$ independently through the convolutional layers in the output side $\{C_{0,j}\}_{j=2}^{I}$.
	This process can alleviate the interference between outputs, while slightly increase the computational costs.
	(2) The work of \cite{zhou2018unet++} manually selects one of the 4 different-level UNet-based structures in the test stage, based on the requirement for speed and accuracy.
	Here, we incorporate IQAM into RBQE and provide early exits in the test stage.
	Therefore, all UNet-based structures are progressively and automatically selected to generate the output.
	The early-exit strategy and proposed IQAM are presented in the following.

	
	\subsubsection{Early-exit Strategy.}
	
	Now we explain the early-exit strategy of RBQE.
	Similarly, we take the RBQE structure with 6 levels as an example.
	The backbone of RBQE can be ablated progressively into 5 different-level UNet-based structures, as depicted in the right half of Fig.~\ref{fig:RBQE}.
	For example, the smallest UNet-based structure with 2 levels consists of 3 convolutional layers: $C_{1,1}$, $C_{2,1}$ and $C_{1,2}$.
	In addition to these 3 layers, the 3-level UNet-based structure includes 3 more convolutional layers: $C_{3,1}$, $C_{2,2}$ and $C_{1,3}$.
	Similarly, we can identify the layers of the remaining 3 UNet-based structures.
	Note that the interval activation layers are omitted for simplicity.
	We denote the parameters of the $i$-level UNet-based structure by $\theta_{i}$.
	This way, the output enhanced samples $\{\mathbf{S}_{\text{out},j}\}_{j=2}^{6}$ can be formulated as:
	\begin{eqnarray}
	\mathbf{S}_{\text{out},j} = \mathbf{S}_{\text{in}} + \mathbf{R}_{j}(\theta_{j}),\ j = 2,3,...,6.
	\end{eqnarray}
	In the test stage, $\{\mathbf{S}_{\text{out},j}\}_{j=2}^{6}$ are obtained and assessed progressively.
	That is, we first obtain $\mathbf{S}_{\text{out},2}$ and send it to IQAM.
	If $\mathbf{S}_{\text{out},2}$ is assessed to be qualified as the output, the quality enhancement process is terminated.
	Otherwise, we further obtain $\mathbf{S}_{\text{out},3}$ and assess its quality through IQAM.
	The same procedure applies to $\mathbf{S}_{\text{out},4}$ and $\mathbf{S}_{\text{out},5}$.
	If $\{\mathbf{S}_{\text{out},j}\}_{j=2}^{5}$ are all rejected by IQAM, $\mathbf{S}_{\text{out},6}$ is output without assessment.
	This way, we successfully perform the early-exit strategy for ``easy'' samples, which are expected to output in the early stage.
	
	\subsection{Image Quality Assessment for Enhanced Images}
	\label{sec:IQAM}

	In this section, we introduce IQAM for blind quality assessment and automatic early-exit decision.
	Most existing blind denoising approaches (e.g., \cite{zhang2017beyond,wang2017novel,yang2018enhancing,guo2019toward}) ignore the characteristics of compression artifacts; however, these characteristics are important to assess the compression artifacts.
	Motivated by \cite{li2016no}, this paper considers two dominant factors that degrade the quality of enhanced compressed images: (1) blurring in the textured area and (2) blocky effects in the smooth area.
	
	Specifically, the enhanced image is first partitioned into non-overlapping patches.
	The patches should cover all potential compression block boundaries.
	Then, these patches are classified into smooth and textured ones according to their sum of squared non-DC Tchebichef moment (SSTM) values that measure the patch energy~\cite{mukundan2001image,li2016no}.
	We take a $4 \times 4$ patch as an example, of which Tchebichef moments can be denoted by $\mathbf{M}$:
	\begin{eqnarray}
		\mathbf{M} = \begin{pmatrix}
			m_{00} & \cdots & m_{03} \\
			\vdots & \ddots & \vdots \\
			m_{30} & \cdots & m_{33}
		\end{pmatrix}.
	\end{eqnarray}
	If the patch is classified as a smooth one, we evaluate its score of blocky effects $\mathcal{Q}_\text{S}$ by calculating the ratio of the summed absolute 3rd order moments to the SSTM value~\cite{li2013referenceless}:
	\begin{eqnarray}
		e_\text{h} = \frac{\sum_{i=0}^3 \vert m_{i3} \vert}{\left( \sum_{i=0}^3 \sum_{j=0}^3 \vert m_{ij} \vert \right) - \vert m_{00} \vert + C},\\
		e_\text{v} = \frac{\sum_{j=0}^3 \vert m_{3j} \vert}{\left( \sum_{i=0}^3 \sum_{j=0}^3 \vert m_{ij} \vert \right) - \vert m_{00} \vert + C},\\
		\mathcal{Q}_{\text{S}} = \log_{(1 - T_{e})} \left(1 - \frac{e_\text{v} + e_\text{h}}{2} \right),
	\end{eqnarray}
	where $e_\text{v}$ and $e_\text{h}$ measure the energy of vertical and horizontal blocky effects, respectively; $C$ is a small constant to ensure numerical stability; $T_{e}$ is a perception threshold.
	The average quality score of all smooth patches is denoted by $\bar{\mathcal{Q}}_\textrm{S}$.
	If the patch is classified as a textured one, we first blur it using a Gaussian filter.
	Similarly, we obtain the Tchebichef moments of this blurred patch $\mathbf{M}'$.
	Then, we evaluate its blurring score $\mathcal{Q}_\text{T}$ by calculating the similarity between $\mathbf{M}$ and $\mathbf{M}'$:
	\begin{eqnarray}
		\mathbf{S}(i,j) = \frac{2 m_{ij} m'_{ij} + C}{(m_{ij})^2 + (m'_{ij})^2 + C},\ i,j = 0,1,2,3,\\
		\mathcal{Q}_{\text{T}} = 1 - \frac{1}{3 \times 3} \sum_{i=0}^3 \sum_{j=0}^3 \mathbf{S}(i,j),
	\end{eqnarray}
	where $\mathbf{S}(i,j)$ denotes the similarity between two moment matrices.
	The average quality score of all textured patches is denoted by $\bar{\mathcal{Q}}_\textrm{T}$.
	The final quality score  $\mathcal{Q}$ of the enhanced image is calculated as
	\begin{eqnarray}
		\mathcal{Q} = (\bar{\mathcal{Q}}_\textrm{S})^{\alpha} \cdot (\bar{\mathcal{Q}}_\textrm{T})^{\beta},
	\end{eqnarray}		
	where $\alpha$ and $\beta$ are the exponents balancing the relative importance between blurring and blocky effects.
	If $\mathcal{Q}$ exceeds a threshold $T_{\mathcal{Q}}$, the enhanced image is directly output at early exits of the enhancement process.
	Otherwise, the compressed image needs to be further enhanced by RBQE.
	Please refer to the supplementary material for additional details.
	
	The advantages of IQAM are as follow:
	(1) IQAM is constructed based on Tchebichef moments~\cite{mukundan2001image}, which are highly interpretable for evaluating blurring and blocky effects.
	(2) The quality score $\mathcal{Q}$ obtained by IQAM is positively and highly correlated to the evaluation metrics of objective image quality, e.g., PSNR and structural similarity (SSIM) index. See the supplementary material for the validation of such correlation, which is verified over 1,000 pairs of raw/compressed images.	
	(3) With IQAM, we can balance the tradeoff between enhanced quality and efficiency by simply tuning threshold $T_{\mathcal{Q}}$.
	
	\subsection{Loss Function}
	
	For each output, we minimize the mean-squared error (MSE) between the input compressed image and output enhanced image:
	\begin{eqnarray}
		\mathcal{L}_j (\theta_j) = \Vert \mathbf{S}_{\text{out},j} (\theta_{j}) - \mathbf{S}_{\text{in}} \Vert_2^2,\ j = 2,3,...,I.
	\end{eqnarray}
	Although MSE is known to have limited correlation with the perceptual quality of images~\cite{wang2004image}, it can still yield high accuracy in terms of other metrics, such as PSNR and SSIM~\cite{guo2019toward,8855019}.
	The loss function of our RBQE approach (i.e., $\mathcal{L}_{\text{RBQE}}$) can be formulated as the weighted combination of these MSE losses:
	\begin{eqnarray}
	\mathcal{L}_{\text{RBQE}} = \sum_{j=2}^{I} w_j \cdot \mathcal{L}_j (\theta_j),
	\end{eqnarray}
	where $w_j$ denotes the weight of $\mathcal{L}_j (\theta_j)$.
	By minimizing the loss function, we can obtain the converged RBQE model that simultaneously enhances the quality of input compressed images with different quality in a resource-efficient manner.
	
	\section{Experiments}
	\label{sec:exp}
	
	In this section, we present the experimental results to verify the performance of the proposed RBQE approach for resource-efficient blind quality enhancement.
	Since HEVC-MSP~\cite{sullivan2012overview} is a state-of-the-art image codec and JPEG~\cite{125072} is a widely used image codec, our experiments mainly focus on quality enhancement of both HEVC-MSP and JPEG images.
	
	\subsection{Dataset}
	
	The recent works have adopted large-scale image datasets such as BSDS500~\cite{amfm_pami2011} and ImageNet~\cite{deng2009imagenet}, which are widely used for image denoising, segmentation and other vision tasks.
	However, the images of these datasets are compressed by unknown codecs and compression settings, thus containing various unknown artifacts.
	To obtain ``clean'' data without any unknown artifact, we adopt the RAISE dataset, from which 3,000, 1,000 and 1,000 non-overlapping raw images are as the training, validation and test sets, respectively.
	These images are all center-cropped into $512 \times 512$ images.
	Then, we compress the cropped raw images by HEVC-MSP using HM16.5 under intra-coding configuration~\cite{sullivan2012overview}, with QP $= 22$, $27$, $32$, $37$ and $42$.
	Note that QPs ranging from 22 to 42 can reflect the dramatically varying quality of compressed images, also in accordance with existing works~\cite{wang2017novel,yang2018enhancing,8855019}.
	For JPEG, we use the JPEG encoder of Python Imaging Library (PIL)~\cite{PIL} to compress the cropped raw images with quality factor (QF) $=10$, $20$, $30$, $40$ and $50$.
	Note that these QFs are also used in \cite{zhang2017beyond}.
	
	\subsection{Implementation Details}
	
	We set the number of levels $I = 6$ for the DNN architecture of RBQE.
	Then, $\{C_{i,1}\}_{i=1}^6$ are conducted by two successive $32 \times 3 \times 3$ convolutions.
	The other $C_{i,j}$ are conducted by two successive separable convolutions~\cite{chollet2017xception}.
	Note that each separable convolution consists of a depth-wise $k \times 3 \times 3$ convolution ($k$ is the input channel number) and a point-wise $32 \times 1 \times 1$ convolution.
	The down-sampling is achieved through a $32 \times 3 \times 3$ convolution with the stride of 2, while the up-sampling is achieved through a transposed $32 \times 2 \times 2$ convolution with the stride of 2.
	For each group of feature maps $\mathbf{F}_{i,j}$, it is further processed by an efficient channel attention layer~\cite{wang2019eca} before being feeding into other convolutional layers.
	Additionally, ReLU~\cite{nair2010rectified} nonlinearity activation is adopted between neighboring convolutions, except the successive depth-wise and point-wise convolutions within each separable convolution.
	For IQAM, we set $\alpha = 0.9$, $\beta = 0.1$, $C=1e{-8}$ and $T_{e} = 0.05$ through a 1000-image validation.
	Additionally, as discussed in Sec.~\ref{sec:comparison}, $T_\mathcal{Q}$ is set to $0.89$ and $0.74$ for HEVC-MSP-compressed and JPEG-compressed images, respectively.
	
	In the training stage, batches with QP from $22$ to $42$ are mixed and randomly sampled.
	In accordance with the ``easy to hard'' paradigm, we set $\{w_j\}_{j=2}^{6}$ to $\{2,1,1,0.5,0.5\}$ for QP $= 22$ or QF $= 50$, to $\{1,2,1,0.5,0.5\}$ for QP $= 27$ or QF $= 40$, to $\{0.5,1,2,1,0.5\}$ for QP $= 32$ or QF $= 30$, to $\{0.5,0.5,1,2,1\}$ for QP $= 37$ or QF $= 20$, and to $\{0.5,0.5,1,1,2\}$ for QP $= 42$ or QF $= 10$.
	This way, high-quality samples are encouraged to output at early exits, while low-quality samples are encouraged to output at late exits.
	We apply the Adam optimizer~\cite{kingma2014adam} with the initial learning rate $\text{lr} = 1e{-4}$ to minimize the loss function.
	
	\subsection{Evaluation}
	\label{sec:comparison}
	
	In this section, we validate the performance of our RBQE approach for the blind quality enhancement of compressed images.
	In our experiments, we compare our approach with 4 state-of-the-art approaches:  DnCNN~\cite{zhang2017beyond}, CBDNet~\cite{guo2019toward}, QE-CNN~\cite{yang2018enhancing} and DCAD~\cite{wang2017novel}.
	Among them, QE-CNN and DCAD are the latest non-blind quality enhancement approaches for compressed images.
	For these non-blind approaches, the training batches of different QPs are mixed and randomly sampled in the training stage, such that they can also manage blind quality enhancement.
	Note that there is no blind approach for quality enhancement of compressed images.
	Thus, the state-of-the-art blind denoisers (i.e., DnCNN and CBDNet) are used for comparison, which are modified for blind quality enhancement by retraining over compressed images.
	For fair comparison, all compared approaches are retrained over our training set.
	
	\subsubsection{Evaluation on efficacy.}
	
	\begin{table}[t]
		\centering
		\scriptsize
		\caption{
			Average $\Delta$PSNR (dB) over the HEVC-MSP and JPEG test sets.
		}
		\vspace{-.3cm}
		\begin{tabular}{c|c c c c c|c|c c c c c}
			\toprule
			\multicolumn{6}{c|}{HEVC-MSP} & \multicolumn{6}{c}{JPEG}\\
			\midrule
			QP & CBDNet & DnCNN & DCAD & QE-CNN & RBQE & QF & CBDNet & DnCNN & DCAD & QE-CNN & RBQE\\
			\midrule
			22 & 0.470 & 0.264 & 0.311 & 0.082 & \textbf{0.604} & 50 & 1.342 & 1.078 & 1.308 & 1.230 & \textbf{1.552}\\
			27 & 0.385 & 0.414 & 0.278 & 0.182 & \textbf{0.487} & 40 & 1.393 & 1.362 & 1.356 & 1.290 & \textbf{1.582}\\
			32 & 0.375 & 0.405 & 0.314 & 0.275 & \textbf{0.464} & 30 & 1.459 & 1.550 & 1.415 & 1.352 & \textbf{1.626}\\
			37 & 0.403 & 0.314 & 0.353 & 0.313 & \textbf{0.494} & 20 & 1.581 & 1.572 & 1.501 & 1.420 & \textbf{1.713}\\
			42 & 0.411 & 0.186 & 0.321 & 0.264 & \textbf{0.504} & 10 & 1.726 & 1.121 & 1.676 & 1.577 & \textbf{1.920}\\	
			\midrule
			ave. & 0.409 & 0.317 & 0.316 & 0.223 & \textbf{0.510} & ave. & 1.500 & 1.337 & 1.451 & 1.374 & \textbf{1.678}\\
			\bottomrule
		\end{tabular}
		\vspace{-.3cm}
		\label{tab:dpsnr_HEVC}
	\end{table}

	\begin{figure}[t]
		\centering
		\subfigure[]{
			\begin{minipage}[h]{0.47\linewidth}
				\centering
				\includegraphics[width=1.0\linewidth]{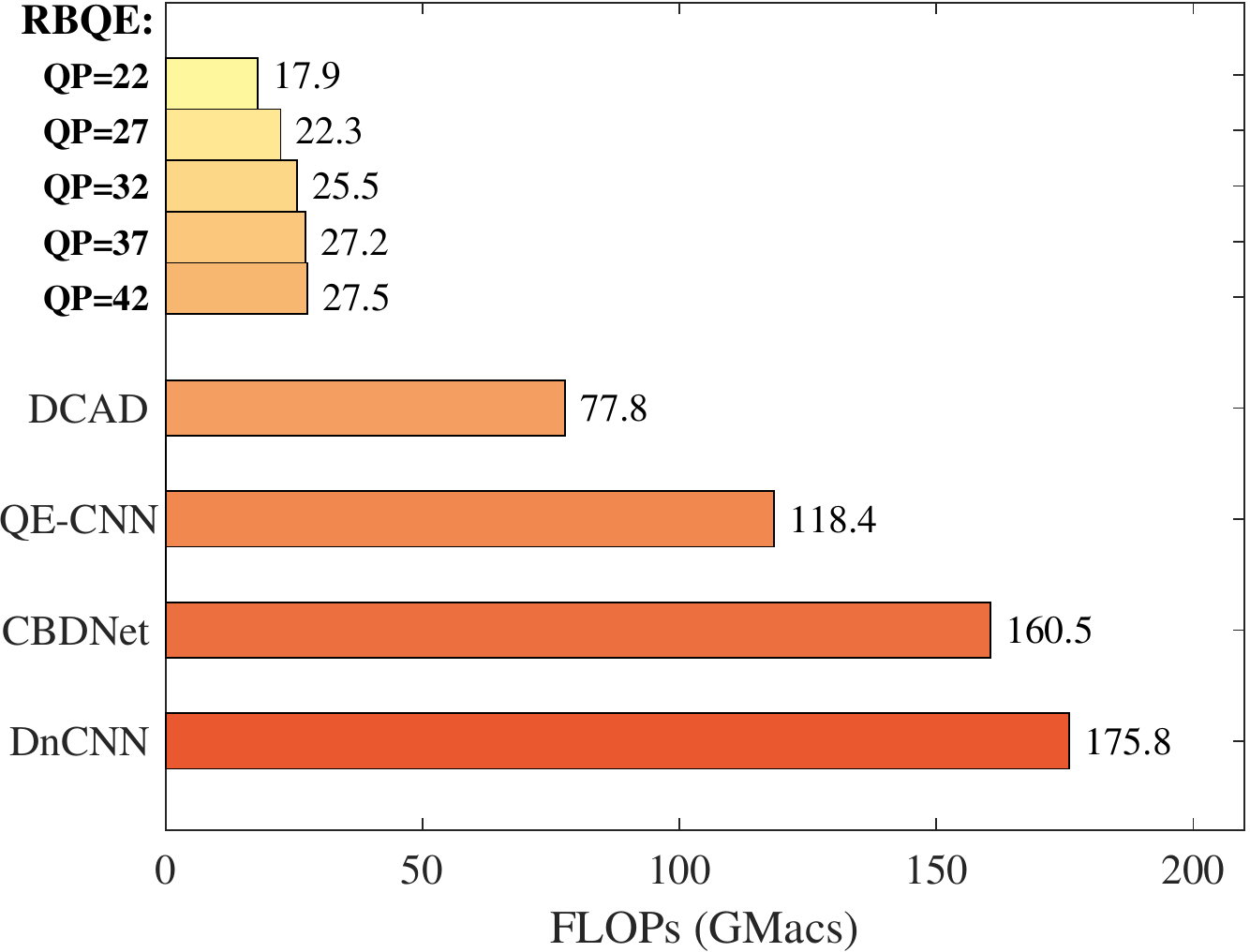}
			\end{minipage}
		}
		\subfigure[]{
			\begin{minipage}[h]{0.47\linewidth}
				\centering
				\includegraphics[width=1.0\linewidth]{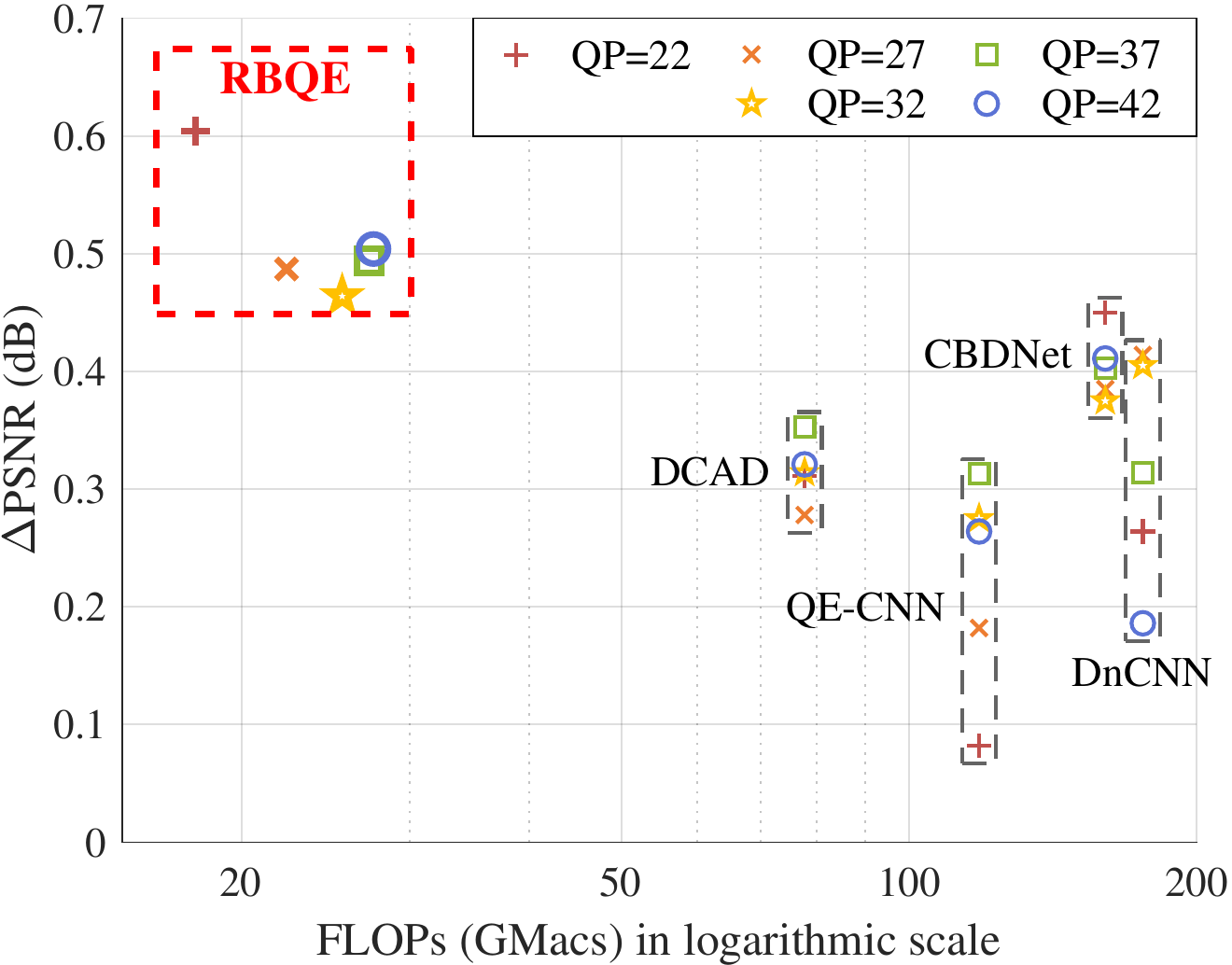}
			\end{minipage}
		}
		\vspace{-.4cm}
		\caption{
			(a) Average FLOPs (GMacs) over the HEVC-MSP test set.
			(b) Average FLOPs (GMacs) vs. improved peak signal-to-noise ratio ($\Delta$PSNR), for blind quality enhancement by our RBQE and compared approaches over the HEVC-MSP test set.
		}
		\vspace{-.3cm}
		\label{fig:FLOPs}
	\end{figure}

	To evaluate the efficacy of our approach, Table~\ref{tab:dpsnr_HEVC} presents the $\Delta$PSNR results of our RBQE approach and other compared approaches over the images compressed by HEVC-MSP.
	As shown in this table, the proposed RBQE approach outperforms all other approaches in terms of $\Delta$PSNR.
	Specifically, the average $\Delta$PSNR of RBQE is 0.510 dB, which is 24.7\% higher than that of the second-best CBDNet (0.409 dB), 60.9\% higher than that of DnCNN (0.317 dB), 61.4\% higher than that of DCAD (0.316 dB), and 128.7\% higher than that of QE-CNN (0.223 dB).
	Similar results can be found in Table~\ref{tab:dpsnr_HEVC} for the quality enhancement of JPEG images.
	
	\subsubsection{Evaluation on efficiency.}
	
	More importantly, the proposed RBQE approach is in a resource-efficient manner.
	To evaluate the efficiency of the RBQE approach, Fig.~\ref{fig:FLOPs} shows the average consumed floating point operations (FLOPs)\footnote{Note that the definition of FLOPs follows \cite{he2016deep,huang2017densely}, i.e., the number of multiply-adds.} by our RBQE and other compared approaches.
	Note that the results of Fig.~\ref{fig:FLOPs} are averaged over all images in our test set.
	As can be seen in this figure, RBQE consumes only 27.5 GMacs for the ``hardest'' samples, i.e., the images compressed at QP $=42$ and 17.9 GMacs for the ``easiest'' samples, i.e., the images compressed at QP $=22$.
	In contrast, DCAD, QE-CNN, CBDNet and DnCNN consume constantly 77.8, 118.4, 160.5 and 175.8 GMacs for all samples that are either ``easy'' or ``hard'' samples compressed at 5 different QPs.
	Similar results can also be found for the JPEG test set, as reported in the supplementary material.
	In summary, our RBQE approach achieves the highest $\Delta$PSNR results, while consuming minimal computational resources especially for ``easy'' samples.
	
	\subsubsection{Tradeoff between efficacy and efficiency.}
	
	\begin{figure}[t]
		\centering
		\subfigure[]{
			\begin{minipage}[h]{0.48\linewidth}
				\centering
				\includegraphics[width=1\linewidth]{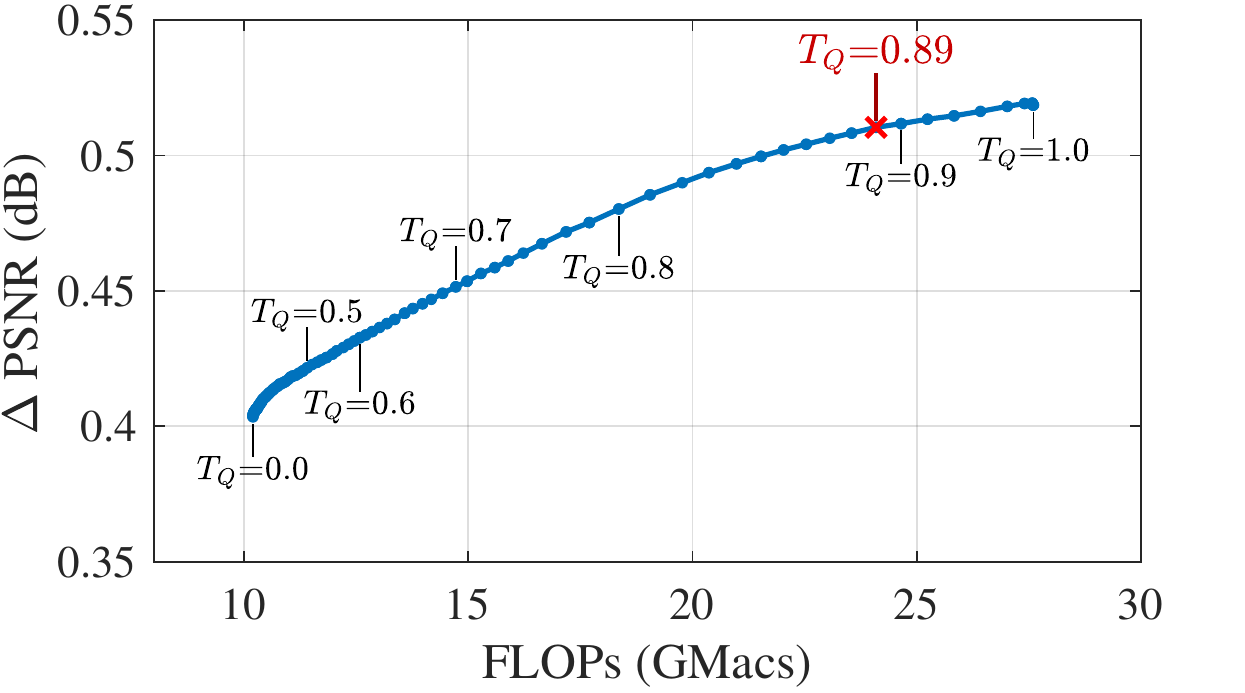}
			\end{minipage}
		}
		\subfigure[]{
			\begin{minipage}[h]{0.47\linewidth}
				\centering
				\includegraphics[width=1\linewidth]{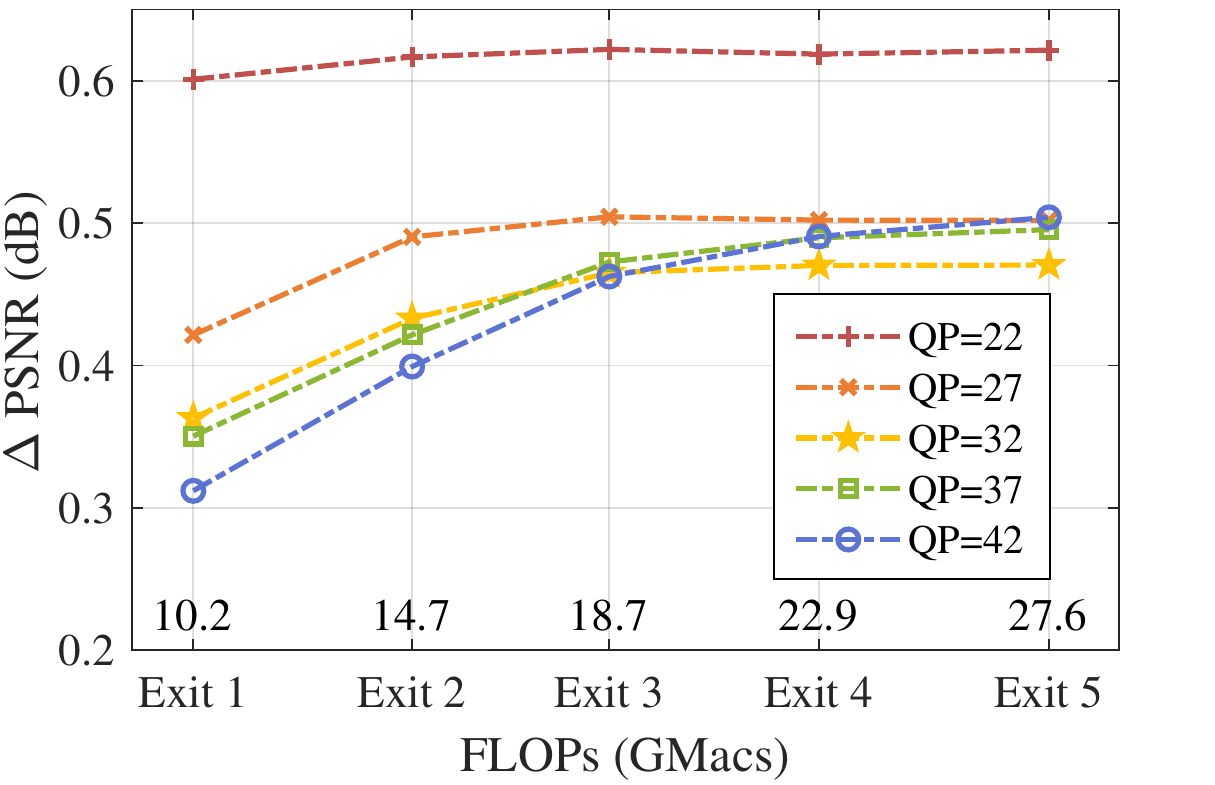}
			\end{minipage}
		}
		\vspace{-.6cm}
		\caption{
			(a) Average $\Delta$PSNR and FLOPs under a series $T_{\mathcal{Q}}$ on HEVC test set.
			(b) Ablation results of the early-exit strategy.
		}
		\vspace{-.5cm}
		\label{fig:tradeoff_HEVC}
	\end{figure}
	
	As aforementioned, we can simply control the tradeoff between efficacy and efficiency by tuning $T_{\mathcal{Q}}$.
	As shown in Fig.~\ref{fig:tradeoff_HEVC} (a), the average $\Delta$ PSNR improves along with the increased consumed FLOPs by enlarging $T_{\mathcal{Q}}$.
	In this paper, we choose $T_{\mathcal{Q}} = 0.89$ for HEVC-MSP-compressed images, since the improvement of average $\Delta$ PSNR gradually becomes saturated, especially when $T_{\mathcal{Q}} > 0.89$.
	Due to the similar reason, we choose $T_{\mathcal{Q}} = 0.74$ for JPEG-compressed images.
	In a word, the tradeoff between efficacy and efficiency of quality enhancement can be easily controlled in our RBQE approach.
	
	\subsubsection{Evaluation on subjective quality.}
	
	\begin{figure}[t]
		\centering
		\includegraphics[width=0.75\linewidth]{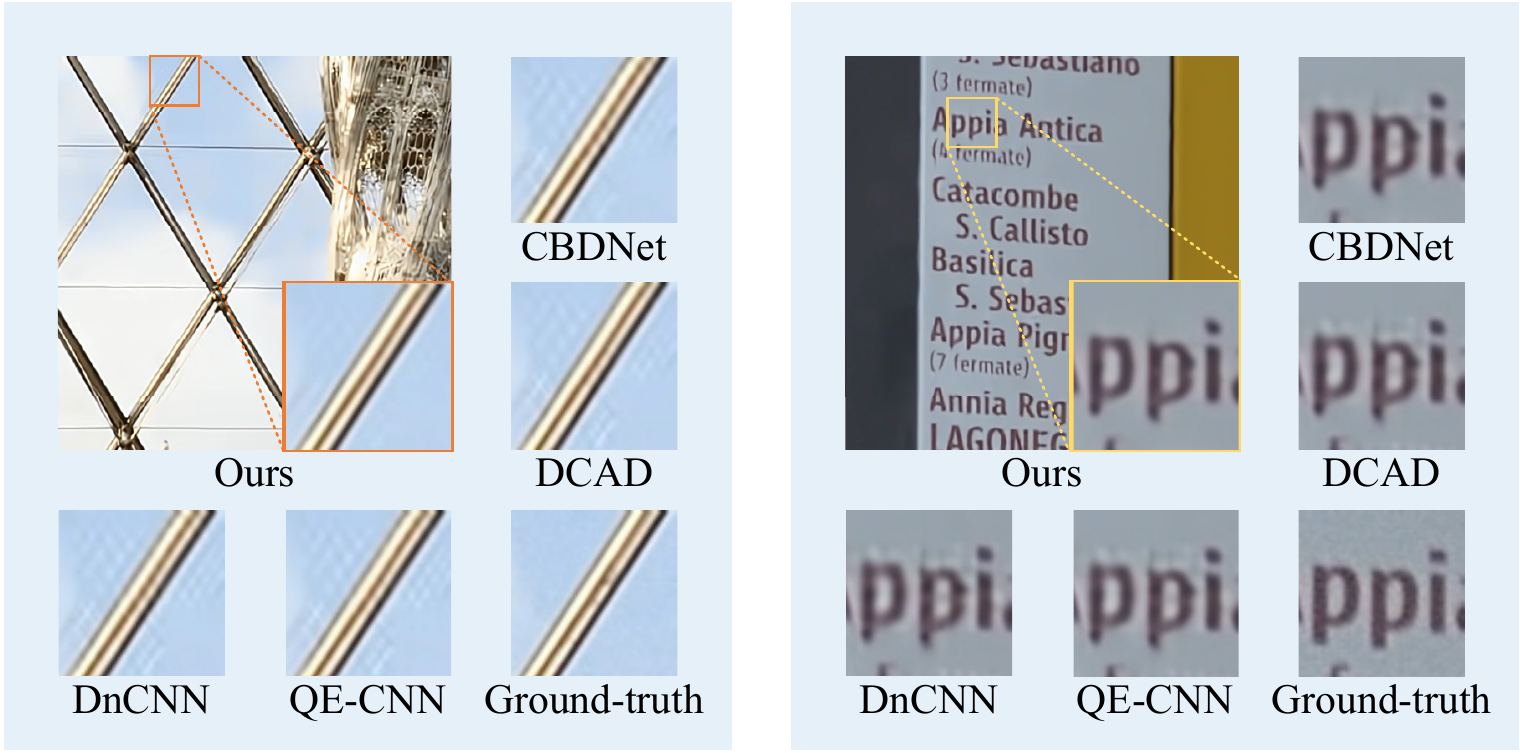}
		\vspace{-.4cm}
		\caption{
			Subjective quality performance on HEVC-MSP test samples.
		}
		\vspace{-.5cm}
		\label{fig:subjective}
	\end{figure}
	
	Fig.~\ref{fig:subjective} visualizes the HEVC-MSP images enhanced by our RBQE and other compared approaches.
	We can observe that our RBQE approach reduces the compression artifacts more effectively than other approaches.
	In particular, the sharp edges can be finely restored by our RBQE approach, while other approaches are ineffective to suppress compression artifacts surrounding the letters and wires.
	More subjective results are presented in the supplementary material.
	
	\subsubsection{Ablation studies.}
	
	To verify the effectiveness of the early-exit structure of our RBQE approach, we progressively ablate the 5 outermost decoding paths.
	Specifically, for the HEVC-MSP images compressed at QP $=22$, we force their enhancement process to be terminated at 5 different exits (i.e., ignoring the automatic decision by IQAM), respectively, and then we obtain the brown curve in Fig.~\ref{fig:tradeoff_HEVC} (b).
	Similarly, we can obtain the other 4 curves.
	As shown in this figure, ``simplest'' (i.e., QP $=22$ ) samples can achieve $\Delta$PSNR $=0.601$ dB at the first exit, which is only $0.02$ dB lower than that at the last exit.
	However, the expense is $270\%$ FLOPs when outputting those samples at the last exit instead of the first one.
	In the opposite, the $\Delta$PSNR of ``hardest'' (i.e.,  QP $=42$) samples output from the last exit is $0.192$ dB higher than that from the first exit.
	Therefore, ``easy'' samples can be simply enhanced while slightly sacrificing quality enhancement performance; meanwhile, more resources provided to ``hard'' samples can result in significantly higher $\Delta$PSNR.
	This is in accordance with our motivation and also demonstrates the effectiveness of the early exits proposed in our RBQE approach.
	
	\section{Conclusions}
	
	In this paper, the RBQE approach has been proposed with a simple yet effective DNN structure to blindly enhance the quality of compressed images in a resource-efficient manner.
	Different from the traditional quality enhancement approaches, the proposed RBQE approach progressively enhances the quality of compressed images, which assesses the enhanced quality and then automatically terminates the enhancement process according to the assessed quality.
	To achieve this, our RBQE approach incorporates the early-exit strategy into a UNet-based structure, such that compressed images can be enhanced in an ``easy to hard'' manner.
	This way, ``easy'' samples can be simply enhanced and output at the early exits, while ``hard'' samples can be further enhanced and output at the late exits.
	Finally, we conducted extensive experiments on enhancing HEVC-compressed and JPEG-compressed images, and the experimental results validated that our proposed RBQE approach consistently outperforms the state-of-the-art quality enhancement approaches, while consuming minimal computational resources.

	%
	%
	\bibliographystyle{splncs04}
	\bibliography{bib}

	\pagestyle{headings}
	\mainmatter
	\def\ECCVSubNumber{2544}  
	
	\title{Supplement to ``Early Exit or Not: Resource-Efficient Blind Quality Enhancement for Compressed Images''} 

	\titlerunning{Supplement to RBQE}
	%
	\author{Qunliang Xing \and
		Mai Xu \and
		Tianyi Li \and
		Zhenyu Guan}
	\authorrunning{Q. Xing et al.}
	%
	\institute{Beihang University, Beijing, China\\
		\email{\{xingql,maixu,tianyili,guanzhenyu\}@buaa.edu.cn}}
	\maketitle
	
	\noindent This material provides the motivation of enhancing JPEG-compressed images in Sec.~\ref{supp:sec:mot}, additional details about image quality assessment module in Sec.~\ref{supp:sec:IQAM}, efficiency performance of enhancing JPEG-compressed images in Sec.~\ref{supp:sec:efficiency}, and more enhanced examples in Sec.~\ref{supp:sec:subjective}.
	
	\section{Motivation of Enhancing JPEG-compressed Images}
	\label{supp:sec:mot}
	
	In this section, we conduct experiments on JPEG-compressed images to prove the two propositions of our paper:
	\textbf{Proposition 1}: ``Easy'' samples (i.e., high-quality compressed images) can be simply enhanced, while ``hard'' samples (i.e., low-quality compressed images) should be further enhanced;
	\textbf{Proposition 2}: The quality enhancement process with different computational complexity can be jointly optimized in a single network through an ``easy to hard'' manner, rather than a ``hard to easy'' manner.
	
	\subsection{Proof of Proposition 1}
	
\begin{figure}[t]
		\subfigure[]{
			\begin{minipage}[h]{0.47\linewidth}
				\centering
				\includegraphics[width=1.0\linewidth]{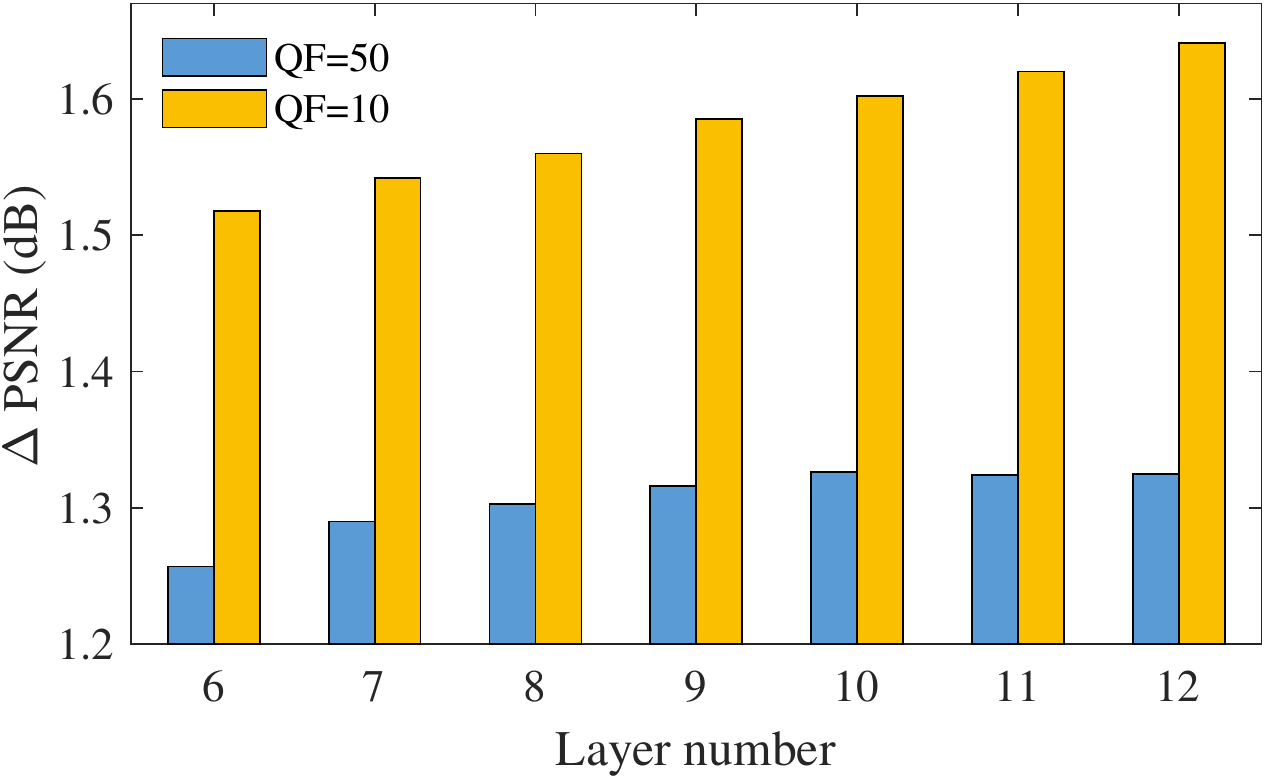}
			\end{minipage}
		}
		\subfigure[]{
			\begin{minipage}[h]{0.47\linewidth}
				\centering
				\includegraphics[width=1.0\linewidth]{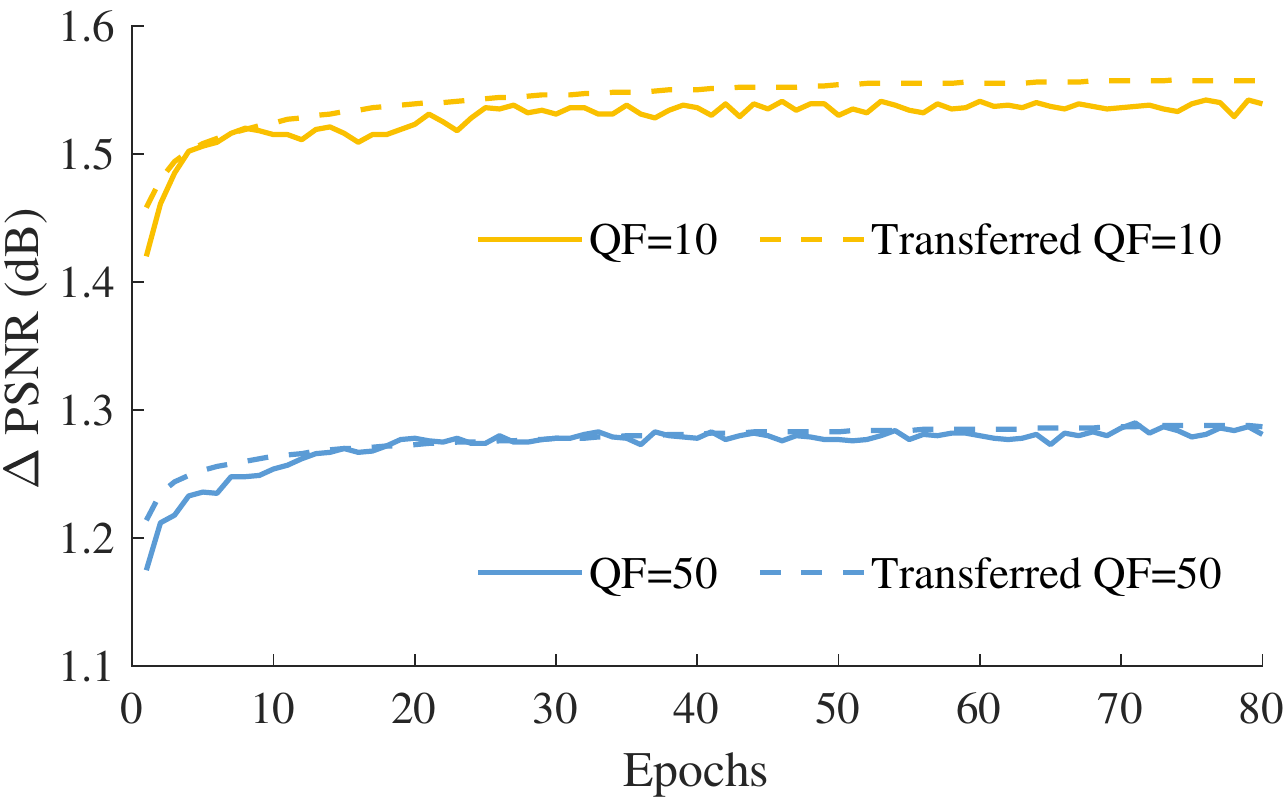}
			\end{minipage}
		}	
		\caption{
			(a) Average $\Delta$PSNR (dB) of vanilla CNNs over the test set.
			(b) Average $\Delta$PSNR (dB) curves alongside increased epochs, for vanilla CNN models and their transferred models over the validation set during the training stage.
		}
		\label{supp:fig:mot}
	\end{figure}

	We construct a series of vanilla CNNs with the layer number from 6 to 12.
	Each layer includes $32 \times 3 \times 3$ filters, except for the last layer with $1 \times 3 \times 3$ filter.
	Beside, ReLU~\cite{nair2010rectified} activation and global residual learning~\cite{he2016deep} are adopted.
	The training, validation and test sets (including 400, 100 and 100 raw images, respectively) are randomly selected from RAISE without overlapping.
	They are all compressed by the JPEG encoder of Python Imaging Library (PIL)~\cite{PIL} with quality factor (QF) $=50$ and $10$ for obtaining ``easy'' and ``hard'' samples, respectively.
	We train these vanilla CNNs with the ``easy'' samples, and then obtain converged models ``QF $=50$''.
	Similarly, we train the CNNs with the ``hard'' samples and then obtain converged models ``QF $=10$''.
	As shown in Fig.~\ref{supp:fig:mot} (a), the performance of QF $=10$ models improves significantly with the increase of layer numbers, while the performance of QF $=50$ models gradually becomes saturated once the layer number excesses 10.
	Therefore, it is possible to enhance the ``easy'' samples with a simpler architecture and fewer computational resources, and to further enhance the ``hard'' samples in a more elaborate process.

	\subsection{Proof of Proposition 2}
	
	Here, we investigate the efficacy of ``easy to hard'' strategy on image quality enhancement through the experiments of transfer learning.
	If the filters learned from ``easy'' samples can be transferred to enhance ``hard'' samples more successfully than the opposite manner, then our proposition can be proved.
	Here, we construct 2 identical vanilla CNNs with 7 convolutional layers.
	The other settings conform to the above.
	We train these 2 models with the training sets of images compressed at QF $=50$ and $10$, respectively, and accordingly these 2 models are called ``QF $=50$'' and ``QF $=10$''.
	After convergence, they exchange their parameters for the first 2 layers and restart training with their own training sets.
	Note that the exchanged parameters are frozen during the training stage.
	We name the model transferred from QF $=10$ to QF $=50$ as ``transferred QF $=50$'' and the model transferred from QF $=50$ to QF $=10$ as ``transferred QF $=10$''.
	Fig.~\ref{supp:fig:mot} (b) shows the validation-epoch curves of the original 2 models and their transferred models.
	As shown in this figure, the transferred QF $=10$ model improves the performance of the QF $=10$ model, while the transferred QF $=50$ model does not benefit the performance of the QF $=50$ model.
	Consequently, the joint simple and elaborate enhancement process should be conducted in an ``easy to hard'' manner rather than a ``hard to easy'' manner.
	In summary, proposition 2 can be proved.
	
	\section{Details about image quality assessment module}
	\label{supp:sec:IQAM}
	
	In this section, we present details about the mechanism of Image Quality Assessment Module (IQAM).
	
	\subsection{Image Partition}
	
\begin{figure}[t]
		\begin{center}
			\subfigure[]{
				\begin{minipage}{0.45\linewidth}
					\centering
					\includegraphics[width=1.0\linewidth]{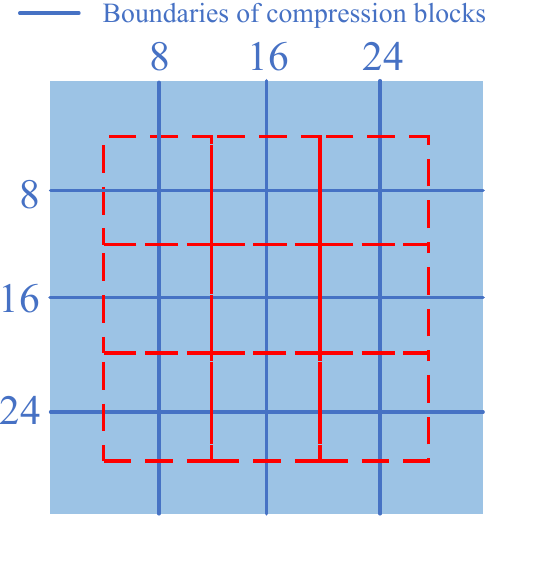}
				\end{minipage}
			}
			\subfigure[]{
				\begin{minipage}{0.45\linewidth}
					\centering
					\includegraphics[width=1.0\linewidth]{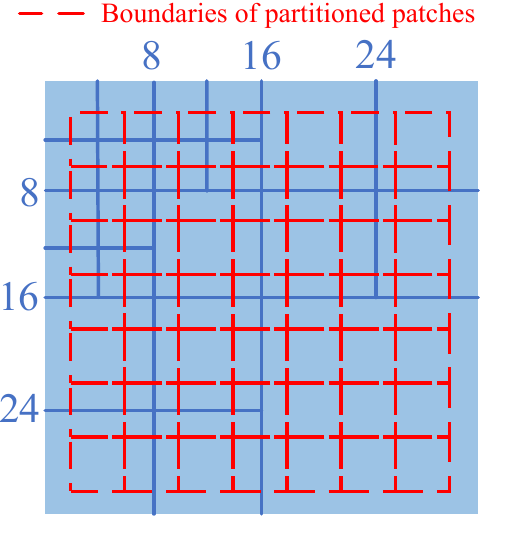}
				\end{minipage}
			}
		\end{center}
		\caption{
			Examples of compression blocks (with blue solid boundaries) and obtained patches (with red dotted boundaries) of a JPEG-compressed image (a) and an HEVC-MSP-compressed image (b).
		}
		\label{supp:fig:patches}
	\end{figure}

	We first partition the image into patches.
	The obtained patches should cover all the potential block boundaries, as shown in Fig.~\ref{supp:fig:patches}.
	For a JPEG-compressed image, it is split into $8 \times 8$ blocks from the top-left.
	Therefore, $8 \times 8$ patches are obtained to cover all block boundaries, as depicted in Fig.~\ref{supp:fig:patches} (a).
	For an HEVC-MSP-compressed image, it is split into coding units (CUs), transform units (TUs) and prediction units (PUs) in a tree structure~\cite{sullivan2012overview}.
	Among these units, CUs and TUs may contribute to blocky effects, and their minimum size is $4 \times 4$.
	Therefore, we partition the HEVC-MSP-compressed image into $4 \times 4$ patches, as depicted in Fig.~\ref{supp:fig:patches} (b).
	
	\subsection{Patches Classification}
	We take $4 \times 4$ patches as an example.
	We classify these patches into smooth ones and textured ones, according to their sum of squared non-DC Tchebichef moment (SSTM) values~\cite{mukundan2001image,li2016no}:
	\begin{eqnarray}
	\mathbf{M} = \begin{pmatrix}
	m_{00} & m_{01} & m_{02} & m_{03} \\
	m_{10} & m_{11} & m_{12} & m_{13} \\
	m_{20} & m_{21} & m_{22} & m_{23} \\
	m_{30} & m_{31} & m_{32} & m_{33}
	\end{pmatrix},
	\end{eqnarray}
	\begin{eqnarray}
	\text{SSTM} = \left( \sum_{i=0}^3 \sum_{j=0}^3 m_{ij}^2 \right) - m_{00}^2,
	\end{eqnarray}
	where $\mathbf{M}$ denotes the Tchebichef moment of each patch.
	In fact, SSTM can be used for measuring block energy, and it is higher for textured patches and lower for smooth patches.
	Therefore, patches can be classified by comparing SSTM with a threshold $T_{\text{SSTM}}$.
	If $\text{SSTM} < T_{\text{SSTM}}$, the patch is classified as a smooth one.
	Otherwise, it will be classified as a textured one.
	
	\subsection{Evaluation on Blocky Effects}
	
	For smooth patches, blocky effects are the dominant factor that degrades the quality.
	Therefore, we evaluate the score of blocky effects in the textured patches.
	Motivated by~\cite{li2013referenceless}, the ratio of the summation of absolute 3th order moment values to the summation of absolute non-DC moment values can reflect the energy of vertical and horizontal blocky effects.
	Specifically, the energy of vertical and horizontal blocky effects, denoted by $e_\text{v}$ and $e_\text{h}$, can be reflected by the following two metrics respectively:
	\begin{eqnarray}
	e_\text{h} = \frac{\sum_{i=0}^3 \vert m_{i3} \vert}{\left( \sum_{i=0}^3 \sum_{j=0}^3 \vert m_{ij} \vert \right) - \vert m_{00} \vert + C},
	\end{eqnarray}
	\begin{eqnarray}
	e_\text{v} = \frac{\sum_{j=0}^3 \vert m_{3j} \vert}{\left( \sum_{i=0}^3 \sum_{j=0}^3 \vert m_{ij} \vert \right) - \vert m_{00} \vert + C},
	\end{eqnarray}
	where $C$ is a small constant to ensure the numerical stability.
	The bigger $e_\text{v}$/$e_\text{h}$, the slighter the vertical/horizontal blocky effects.
	When $e_\text{v}$/$e_\text{h}$ is bigger than a threshold $T_{e}$, the blocky effects are too slight to notice:
	\begin{eqnarray}
	e_\text{v}=\left\{
	\begin{array}{rcl}
	e_\text{v} & {e_\text{v} < T_{e}}\\
	T_{e}      & {e_\text{v} \geq T_{e}},
	\end{array} \right.
	\end{eqnarray}	
	\begin{eqnarray}
	e_\text{h}=\left\{
	\begin{array}{rcl}
	e_\text{h} & {e_\text{h} < T_{e}}\\
	T_{e} & {e_\text{h} \geq T_{e}},
	\end{array} \right.
	\end{eqnarray}
	Finally, the quality score of the smooth patch is calculated as:
	\begin{eqnarray}
	\mathcal{Q}_{\text{S}} = \log_{(1 - T_{e})} \left(1 - \frac{e_\text{v} + e_\text{h}}{2} \right).
	\end{eqnarray}
	Note that a higher $\mathcal{Q}_{\text{S}}$ indicates better quality of the smooth patch.
	
	\subsection{Evaluation on Blurring}
	
	For textured patches, blurring is the dominant factor that degrades the quality.
	Therefore, we evaluate the score of blurring in the textured patches.
	Each textured patch is blurred by a Gaussian filter:
	\begin{eqnarray}
	G(x,y,\sigma) = \frac{1}{2 \pi \sigma^2} \exp \left( \frac{- (x^2 + y^2)}{2 \sigma^2} \right),
	\end{eqnarray}
	where $\sigma$ is the stand deviation of the Gaussian filter; $(x,y)$ denotes the coordinate of each image pixel.
	Following \cite{li2016no}, the filter size is set to $3 \times 3$.
	The standard deviation is set to $5$.
	Then, we can obtain the Tchebichef moments of the blurred patch:
	\begin{eqnarray}
	\mathbf{M'} = \begin{pmatrix}
	m'_{00} & m'_{01} & m'_{02} & m'_{03} \\
	m'_{10} & m'_{11} & m'_{12} & m'_{13} \\
	m'_{20} & m'_{21} & m'_{22} & m'_{23} \\
	m'_{30} & m'_{31} & m'_{32} & m'_{33}
	\end{pmatrix}.
	\end{eqnarray}
	The similarity of $\mathbf{M}$ and $\mathbf{M'}$ is calculated as:
	\begin{eqnarray}
	\mathbf{S}(i,j) = \frac{2 m_{ij} m'_{ij} + C}{(m_{ij})^2 + (m'_{ij})^2 + C},\ i,j = 0,1,2,3.
	\end{eqnarray}
	The similarity value of severely blurred patch is bigger than that of slightly blurred patch.
	Therefore, we compute the quality score of textured patch as follows,
	\begin{eqnarray}
	\mathcal{Q}_{\text{T}} = 1 - \frac{1}{3 \times 3} \sum_{i=0}^3 \sum_{j=0}^3 \mathbf{S}(i,j).
	\end{eqnarray}
	Note that a higher $\mathcal{Q}_{\text{T}}$ indicates better quality of the textured patch.
	Also note that this formulation of quality score of textured patch is different from that in \cite{li2016no}, which directly takes the average similarity values as the quality score.
	It is inconsistent with the quality score of smooth patch, since the latter increases when the quality is better.
	
	\subsection{Quality Score}
	
	Finally, we calculate the average quality scores of all smooth patches:
	\begin{eqnarray}
	\bar{\mathcal{Q}}_{\text{S}} = \sum_{k=1}^{N_\text{S}} \mathcal{Q}_{\text{S,k}},
	\end{eqnarray}	
	where $\mathcal{Q}_{\text{S,k}}$ is the $k$-th smooth patch and $N_\text{S}$ is the number of smooth patches.
	Similarly, we can obtain the average quality scores of all textured patches $\bar{\mathcal{Q}_{\text{T}}}$.
	The final quality score of the image can be generated as:
	\begin{eqnarray}
	\mathcal{Q} = (\bar{\mathcal{Q}}_\text{S})^{\alpha} \cdot (\bar{\mathcal{Q}}_\text{T})^{\beta},
	\end{eqnarray}	
	where $\alpha > \beta$.
	It is because blocky effects are the dominant distortion~\cite{li2016no}, especially for low bit-rate compression.
	
	\subsection{Implementation}
	
	We set $T_{\text{SSTM}} = 4e{-3}$, $C=1e{-8}$, $T_{e} = 5e{-2}$ and $\alpha = 0.9 > \beta = 0.1$ through experiments on a validation set (including 1000 pairs of raw/compressed images randomly selected from RAISE~\cite{dang2015raise}) compressed at 5 different QPs (QP $=22,27,32,37,42$).
	
	\subsection{Relation between $T_{\mathcal{Q}}$ and Objective Quality Metrics}
	
\begin{figure}[t]
		\begin{center}
			\subfigure[]{
				\begin{minipage}{0.47\linewidth}
					\centering
					\includegraphics[width=1\linewidth]{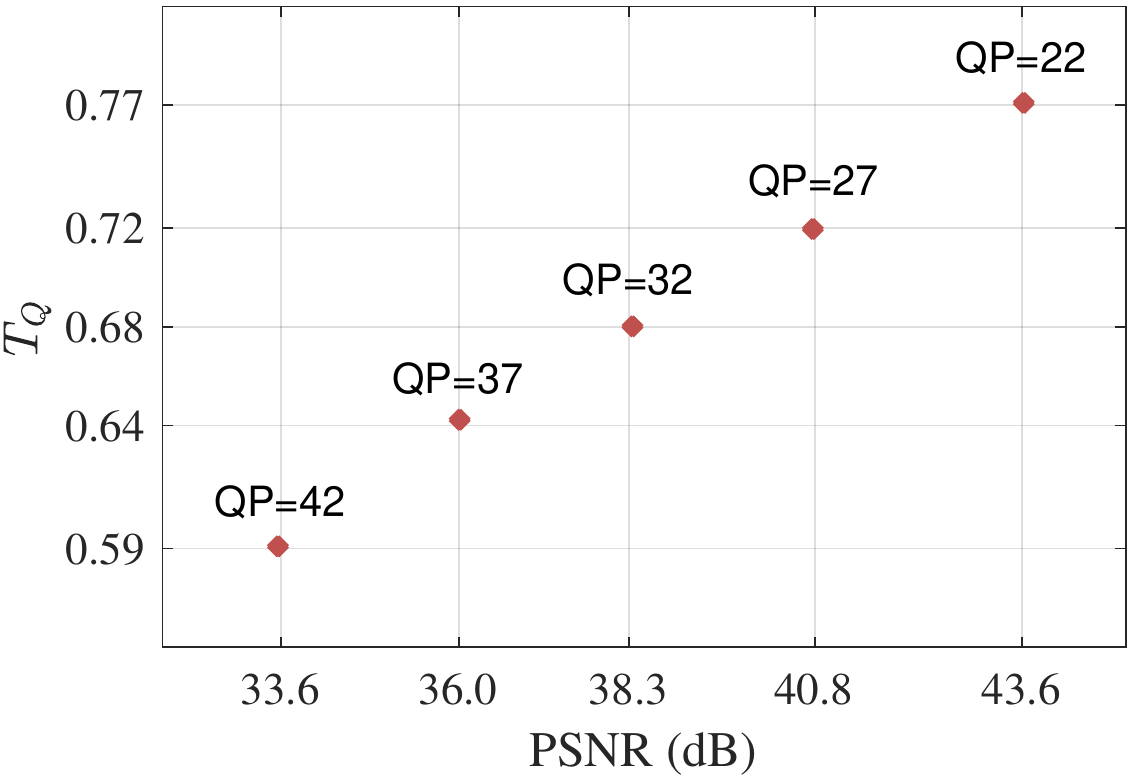}
				\end{minipage}
			}
			\subfigure[]{
				\begin{minipage}{0.47\linewidth}
					\centering
					\includegraphics[width=1\linewidth]{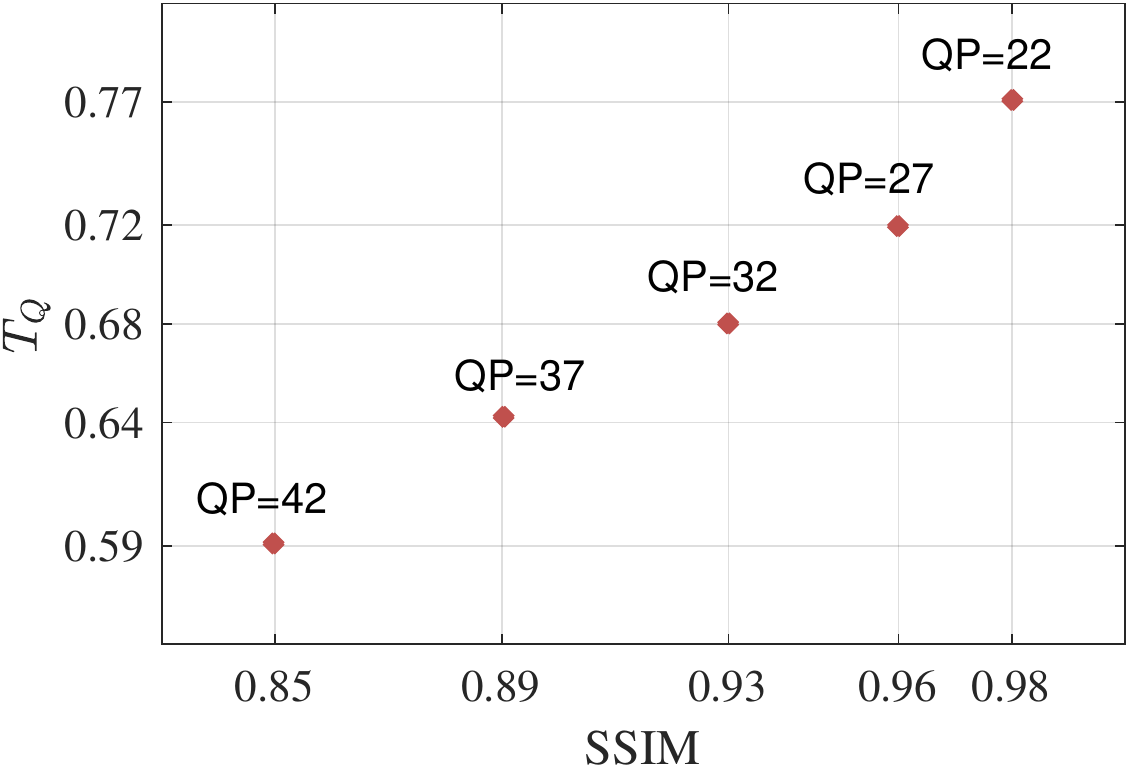}
				\end{minipage}
			}
		\end{center}
		\caption{
			(a) Average PSNR (dB) and $T_{\mathcal{Q}}$ of the HEVC validation set compressed at 5 QPs.
			(b) Average SSIM and $T_{\mathcal{Q}}$ of the HEVC validation set compressed at 5 QPs.
		}
		\label{supp:fig:corr}
	\end{figure}

	The generated quality score $T_{\mathcal{Q}}$ is highly and positively correlated to objective quality metrics, such as peak signal-to-noise ratio (PSNR) and structural similarity (SSIM) index.
	To verify this, we calculate the average $T_{\mathcal{Q}}$ of the validation set.
	As shown in Fig.~\ref{supp:fig:corr} (a) and (b), the PSNR, SSIM and $T_{\mathcal{Q}}$ increase along with the decreased QP values.
	Therefore, the strong and positive correlation between $T_{\mathcal{Q}}$ and objective quality metrics PSNR and SSIM can be verified.

	\section{Efficiency of Enhancing JPEG-compressed Images}
	\label{supp:sec:efficiency}
	
	We also validate the efficiency of the RBQE approach when enhancing the quality of JPEG-compressed images in terms of the average consumed FLOPs.
	Our RBQE approach consumes only 26.9 GMacs for the ``hardest'' samples, i.e., the images compressed at QF $=10$.
	In contrast, DCAD, QE-CNN, CBDNet and DnCNN consume constantly 77.8, 118.4, 160.5 and 175.8 GMacs for all samples that are either ``easy'' or ``hard'' samples compressed at 5 different QFs.
	Therefore, our RBQE approach is much more efficient than compared approaches when enhancing the quality of JPEG-compressed images.
	
	\section{Additional Enhanced Samples}
	\label{supp:sec:subjective}
	
	\begin{figure}[t]
		\begin{center}
			\includegraphics[width=1.0\linewidth]{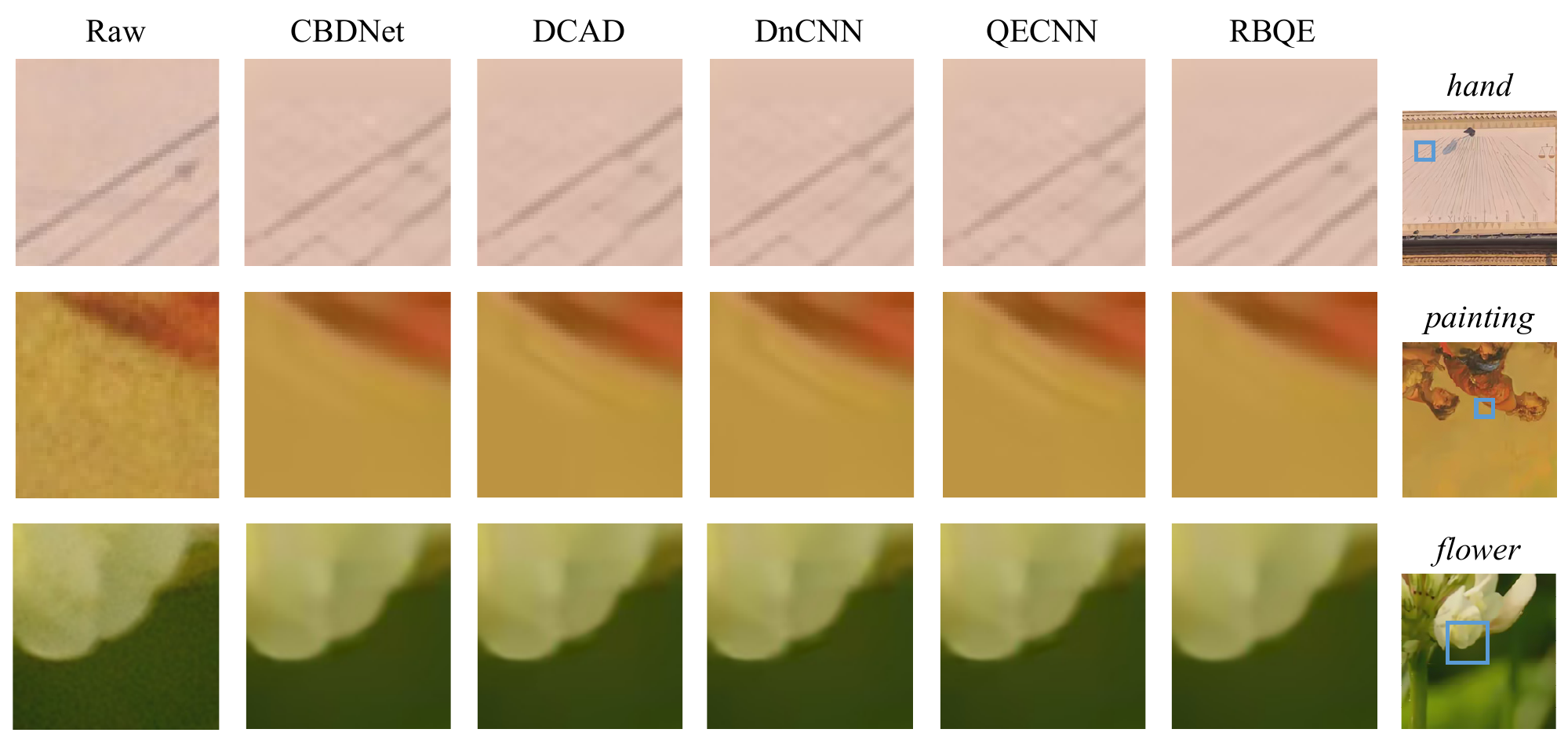}
		\end{center}
		\caption{
			Enhanced test samples.
			We observe significant suppression by RBQE of compression artifacts surrounding the lines, clothes and petals.
		}
		\label{supp:fig:subjective}
	\end{figure}
	
	Fig.~\ref{supp:fig:subjective} visualizes the enhanced samples by our RBQE and other compared approaches.
	In particular, the smooth background can be finely restored by our RBQE approach, while other approaches are ineffective to suppress compression artifacts surrounding the lines, clothes and petals.

	%
	%
\end{document}